\documentclass{aa}  

\usepackage{graphicx}
\usepackage{txfonts}
\usepackage{subcaption}         
\usepackage{lscape}             
\usepackage{placeins}           

\usepackage[normalem]{ulem} 
\usepackage{amsfonts} 
\usepackage[dvipsnames]{xcolor}
\usepackage{eqnarray}
\usepackage{wrapfig} 
\usepackage{float} 
\usepackage{mathtools}

\usepackage{colortbl}

\usepackage{hyperref} 
\hypersetup{colorlinks, urlcolor=CadetBlue, citecolor=PineGreen, linkcolor=gray} 

\usepackage{multirow}
\usepackage{natbib} 

\usepackage{amsmath}	
\usepackage{amssymb}	
\usepackage{multicol}        
\usepackage{bm}		

\defcitealias{Ibata2021}{I21}
\defcitealias{Thomas2020}{T20}
\defcitealias{Shipp2019}{S19}

\begin{document}

   \title{Systematic census of RR Lyrae stars in Milky Way stellar streams}



    \author{Bruno Domínguez\inst{1} \thanks{e-mail:
            \href{mailto:bdominguez@fcien.edu.uy}{bdominguez@fcien.edu.uy}}
            \and
            Cecilia Mateu\inst{1} 
            \and
            Pau Ramos\inst{2}
            \and Guillaume F. Thomas \inst{3,4}
            }

   \institute{Departamento de Astronomía, Facultad de Ciencias,     
              Universidad de la República, Iguá 4225, 14000, Montevideo, Uruguay
             \and
             National Astronomical Observatory of Japan, Mitaka-shi, Tokyo 181-8588, Japan
             \and
             Universidad de La Laguna, Av. Astrofísico Francisco Sánchez E-38205 La Laguna, Santa Cruz de Tenerife, España.
             \and 
             Instituto de Astrofísica de Canarias, Calle Vía Láctea s/n E-38206 La Laguna, Santa Cruz de Tenerife, España}

   \date{Received YYY; accepted XXX}

 
  \abstract
   {Nearly 150 tidal streams are known today in the Milky Way; yet, full phase-space information exists for only a handful. RR Lyrae stars (RRL) -- pulsating horizontal branch stars that serve as excellent standard candles -- offer a powerful means to probe these structures, but they have only been specifically identified in less than dozen streams.}
   {We study the RRL population in all the known stellar streams with reported proper motion in the galstreams library, performing the first systematic census of these stars across the streams. Our goals are to identify likely RRL members, map distances along streams, and compare RRL populations in streams with those in their surviving progenitors.}
   {We use a union of the largest RRL catalogs (Gaia~DR3 SOS, PS1, and ASAS-SN-II) to construct a Bayesian probabilistic membership model and find 361 RRL in the 56 streams studied.}
   {\textit{i)} We find that 32 of the 56 streams have RRL in their tidal tails -- 13 with progenitors and 19 without. Of these, 13 streams present more than 3 RRL in their tails. \textit{ii)} We report new RRL detections in 31 of the 32 streams with already identified RRL populations, anchoring the distances of the streams and, in particular, inferring new distance gradients for 5 of them. \textit{iii)} Our method also provides intrinsic dispersion estimates in distance and proper motion for each track and statistically quantifies the expected contamination. \textit{iv)} In addition, the census revealed some complex origin histories, such as the new plausible origin scenario we propose for M92 with multiple progenitors. \textit{v)} We also find that the presence of RRL in the tidal tails is linked to the late stages of progenitor dissolution.}
   {This census represents a first step toward identifying which of the studied stellar streams contain a significant number of RRL based on currently reported tracks while also providing a homogeneous and robust catalog of RRL members with precise empirical distances, crucial for a full phase-space analysis of these structures and their use as probes of the Galaxy's history and gravitational potential.}


   \maketitle
%

\section{Introduction}

Currently, it is known that galaxies like our own have grown by gradually assimilating smaller galaxies in what is called `hierarchical formation' \citep[e.g.,][]{White1978,White1991}. These accretion events have left behind fossil records that can be discovered by studying the most ancient populations, such as the Galactic halo \citep[e.g.,][]{Gallart2019,Naidu2020,Helmi2020,Deason2024}. An important part of these fossil records are the globular clusters present in the halo, many of which were brought into the Milky Way through the accretion of their respective host galaxies \citep[e.g.,][]{Searle1978,Zinn1993,Kruijssen2019,Callingham2022,Bellazzini2022}. But these are only the surviving clusters. During the accretion process, clusters and dwarf galaxies can be disrupted due to tidal forces, causing the stars to escape from the progenitor, approximately tracing its orbit and producing a dynamically coherent structure called `stellar stream' \citep[e.g.,][]{Helmi1999,Combes1999,Eyre2009}.

Thanks to the arrival of large area photometric surveys such as the Sloan Digital Sky Survey \citep[SDSS,][]{SDDS}, the \textit{Gaia} mission \citep{Gaia2016}, the Panoramic Survey Telescope and Rapid Response System~1 \citep[Pan-STARRS1 (PS1),][]{Chambers2016}, the  Ultraviolet Near-infrared Optical Northern Survey \citep[UNIONS,][]{Gwyn2025}, and the Dark Energy Survey \citep[DES,][]{Shipp2018}, almost 150 stellar streams are known in the Galaxy today \citep{Mateu2023}, most of which have likely been produced by globular clusters \citep{Bonaca2025}. Being dynamically cold structures, they are powerful tools for inferring the gravitational potential of the Milky Way (e.g., \citealt{Kupper2015}, \citealt{Ibata2021}, hereafter \citetalias{Ibata2021}; \citealt{Ibata2024}; \citealt{Palau2025}) providing constraints on the enclosed mass and the distribution of dark matter; as well as the nature of dark matter itself through the study of perturbations such as gaps, blobs, and spurs, from which constraints on the underlying distribution of subhalos could be inferred \citep[e.g,][]{Erkal2016,Price-Whelan2018, Bonaca2019b, deBoer2020}. In addition, streams are sensitive to encounters with massive objects like the Magellanic Clouds or the Sagittarius dwarf galaxy, which can cause misalignment between their celestial paths and the velocities of their members \citep[e.g.,][]{Erkal2019,Vasiliev2021,Mateu2023}. Unveiling these interactions can help to decipher the dynamic history of the Milky Way and to rebuild its past. On smaller scales, residuals of past accretion events can also be traced in stellar streams that exhibit multiple components, such as a cocoon feature stellar streams that exhibit multiple components, such as a cocoon feature, can be used to trace the residuals of past accretion events, further enriching the fossil record of the Galaxy assembly \citep{Malhan2021b}. Moreover, studying the dynamic properties of the stellar streams, such as their integrals of motion, also allows us to associate different streams with the different massive accretion events that built our Galaxy and to obtain a more complete picture of its formation history \citep{Bonaca2021}. 
In order to conduct all these kinds of studies, full phase-space information for the stellar streams is required; yet, empirical distance information exists for only about half of the known streams \citep{Mateu2023}.

One way to overcome this issue is to work with RR Lyrae stars (RRL), pulsating horizontal branch giants that serve as excellent standard candles. There are mainly two types of RRL: the \emph{ab} type (RRab), which pulsate in their fundamental mode, and the \emph{c} type (RRc), which pulsate in their first overtone mode \citep{Smith1995}. 
RRL are abundant in globular clusters and have proven to be excellent tracers of the canonically oldest ($\gtrsim$ 10 Gyr) and metal-poor ([Fe/H]$<-0.5$) populations in the Galaxy, mainly the halo \citep{Vivas2006,Vivas2006b,Iorio2021}, bulge \citep{Kunder2008,Prudil2024,Prudil2024b}, and thick disk \citep[][and references therein]{Mateu2018}. Being luminous stars and one of the few astronomical objects for which distances can be measured with great precision ($\lesssim 5\%$), they serve as ideal markers for studying the six-dimensional structure of the Galaxy when combined with all-sky kinematic information \citep{Iorio2021}. Therefore, by identifying which RRL belong to the different stellar streams, we can anchor a precise mean distance to all of them and, in cases with numerous RRL populations, determine a distance gradient along the stream. 
In addition, we can estimate the masses of the streams and study the properties of their population compared to the progenitor population while the latter still exists, allowing us to understand more about the origins of the stellar system.

Despite the powerful utility of the RRL, so far only a handful of streams have been characterized using RRL: Orphan \citep{Koposov2019}, Palomar 5 \citep[Pal 5,][]{Price-Whelan2019}, LMS-1 \citep{Yuan2020}, and Sagittarius \citep[e.g.,][]{Sesar2017,Hernitschek2017,Ramos2020}; however, these studies rely on completely different catalogs and methods -- some with kinematic information and some without -- making comparison of their results difficult. There have also been a few studies that have systematically searched for RRL in stellar streams with associated progenitors: \citet{Kundu2019} and \citet{Abbas2021} in globular clusters and \citet{Vivas2020} and \citet{Tau2024} in dwarf galaxies. However, in these studies, the RRL were visually associated based on positional, proper motion, and color-magnitude diagram criteria. In addition, several other studies (e.g., \citealt{Shipp2019}, hereafter \citetalias{Shipp2019}; \citealt{Hanke2020}; \citetalias{Ibata2021}; \citealt{Xu2024}; \citealt{Li2019}) have searched for stream members in general and, in some cases, have identified RRL as well; however, these works were not specifically designed to target RRL or exploit their properties to study the stream populations. 

These precedents highlight the need to conduct a systematic and robust RRL census of all known streams, which, additionally to enabling a homogeneous study of the RRL stream population, also allows for a comparison between the populations in clusters and in their tidal tails for streams with surviving progenitors particularly, since RRL are very common in globular clusters. For example, the RRL population of the Pal 5 stream was characterized in detail by \citet{Price-Whelan2019} using a probabilistic kinematics membership model. They identified 10 RRL within the tidal radius of the cluster and nearly twice as many in the tidal tails. 
This study revealed an intriguing segregation in the RRL population of the stream: while most of the RRc (67\%) are in the cluster, the majority of the RRab (87\%) are in the tails. This raises several questions: Is such segregation a standard feature across streams or just a peculiarity of Pal 5? Could other types of segregation exist in the stellar population of the stream that are not reported yet? How frequent are RRL in the tidal tails of globular clusters?

The goal of this work is, therefore, to study the RRL population of known stellar streams by performing a systematic census. 
It structure is as follows: In Section \ref{sc:data}, we describe the data used. In Section \ref{sc:method1}, we describe the probabilistic membership model implemented to identify candidate members of the streams. In Section \ref{sc:result}, we present the identified RRL members for each stream individually and discuss their properties. Meanwhile, in Section \ref{sc:discussion} we use our results to discuss the statistical properties of the population of streams and shed light on the possible origins of a selection of streams. Finally, in Section \ref{sc:conclu} we present the conclusions and summarize the contributions of the thesis.

\section{Data} \label{sc:data}

We used  a catalog of 309,998 RRL conformed by the union of the Gaia~DR3 Specific Objects Study \citep[SOS,][]{Clementini2023}, PS1 and All-Sky Automated Survey for Supernovae II \citep[ASAS-SN-II,][]{Jayasinghe2019} RRL catalogs, the largest and most complete RRL catalogs currently available \citep{Mateu2020, Mateu2024}, allowing us to trace structures up to $\sim 100$ kpc. This compilation is similar to the catalog used in \citet{Cabrera2024}, where the details of distance calculations are explained, except that in this work, for stars without a photometric metallicity estimate, we assigned a metallicity drawn from a normal distribution with a mean of $-1.2$~dex and a standard deviation $0.5$~dex mimicking the average properties of (kinematically selected) halo stars in Gaia~DR3 SOS according to \citet{Li2023}. We restricted the final catalog to stars with RUWE $<$ 1.4 to ensure the quality of the astrometric parameters, following the recommendation of \citet{Lindegren2021}.
 
In this paper we are not trying to find new streams, but to find RRL members of those already known. To do this, we use \verb|galstreams| \citep{Mateu2023}, the most complete library of stellar streams in the Milky Way with reported tracks in the sky, proper motion, distance, and radial velocity when available. 
We will follow the same nomenclature introduced in their work: A \emph{stream} is the object of study and each is unique, while a \emph{track} is the measure of a property (sky position, proper motion or radial velocity) of the stream as a function of an angle. Different tracks may have been reported in the literature by different authors; therefore, in many cases multiple tracks exist for the same stream. Full details are explained in \citet{Mateu2023}. It is worth mentioning that \citet{Mateu2023} refers to \verb|galstreams v1.0|, while in this work we use \verb|galstreams v1.2|, whose main updates are: the addition of more than 40 new streams, more than 40 new tracks for previously known streams (including radial velocity tracks), and information about the observed widths in the sky and in proper motion. In this version, there are 217 tracks that correspond to 137 unique streams, of which only 56 streams (75 tracks\footnote{12 of the studied streams have more than one track associated.}) have proper motion information\footnote{The properties used for each track are summarized in Table \ref{tb:resumen_props} in Appendix \ref{a:track_props}.}. We exclude from the present analysis the tracks of Sagittarius-A20 \citep{Antoja2020,Ramos2020, Ibata2020, Ramos2022}, LMS1-Y20 \citep{Yuan2020}, and Cetus-Palca-Y21 \citep{Yuan2022} because these are too long for our method to yield reliable results. We also exclude the Monoceros-R21 and ACS-R21 tracks \citep{Ramos2021} because, even though they are included in \verb|galstreams|, as noted by \citet{Mateu2023} these structures are not considered to be real stellar streams but overdensities formed by stars kicked out of the Galactic plane \citep[e.g.,][]{deBoer2018,Laporte2020}.

In the case of streams with an associated progenitor, we used the astrometric, kinematic, and structural properties of globular clusters from the Database of Fundamental Parameters of Galactic Globular Clusters\footnote{\url{https://people.smp.uq.edu.au/HolgerBaumgardt/globular/}} \citep[DbGC,][]{Baumgardt&Hilker2018, Baumgardt&Vasiliev2021, Vasiliev&Baumgardt2021, Baumgardt2023}. For Tucana III, the only surviving dwarf galaxy with an associated stream in our sample, we used data from \citet{Drlica-Wagner2015}.

\section{Method} \label{sc:method1}

With the aim of associating the RRL with the different stellar streams, we constructed a Bayesian probabilistic membership model. Taking the sky (celestial), distance and proper motion tracks reported in \verb|galstreams| as true, each star is modeled as having a probability of belonging to the stream ($p_{\text{st}}$) in addition to a probability of belonging to the background ($ p_{\text{bkg}}$), both of which depend exclusively on the star's distance and proper motion. We then estimate the membership probability as described below, and if this probability is higher than 50\% and the star falls within the extended stream region in the sky, we classify that RRL as a member of the stream. In  what follows, we describe the sample selection criteria used followed by the details of the Bayesian model.

\subsection{Sample selection} \label{sc:selection}

For each stream, we restrict our search for members to a sky window  defined in the stream frame of reference $\phi_1-\phi_2$, the rotated spherical coordinate system aligned with the stream, as reported in \verb|galstreams|. To define the window, we start by taking the sky track and width ($\sigma_{\phi_2}$) of the stream from \verb|galstreams|. The length in $\phi_1$ is taken as 1.5 times the length of the sky track in $\phi_1$. For the window limits in $\phi_2$, we take the maximum (minimum) of the sky track in $\phi_2$ and add (subtract) 11$\sigma_{\phi_2}$\footnote{The size of the sky window was taken arbitrarily, but large enough to cover the stream and the control areas around it.}, placing the stream at the center of the window in both directions. Within this sky window, we define two regions of interest: an on-stream region defined as a polygonal footprint around the sky track with a width equal to 6$\sigma_{\phi_2}$, and an off-stream region defined as two footprints parallel to the stream footprint, one immediately above (in $\phi_2$) and the other immediately below, each with a width equal to 6$\sigma_{\phi_2}$ and extended in all $\phi_1$ within the sky window.

Finally, we also define limits along the line of sight by making a box for our search: we extrapolate the distance track up to the limits of the sky window in $\phi_1$, select its maximum (minimum) value, and add (subtract) to it 4 times its 10\% which corresponds to the mode of the distance error distribution. 
In Table~\ref{tb:resumen_props} we show a summary of the properties used for each track.

Once the box is defined, we remove from the sample all known globular clusters and dwarf galaxies that happen to fall inside our selection but are unrelated to the stream of interest. We do that by removing the stars within the tidal radius, $r_t$, of each of the contaminant objects. We also remove all RRL with latitude $|b|<15^\circ$ (or $|b<|b_{\text{prog}}|-r_t$ in the case of stream progenitors with latitude $|b_{\text{prog}}|<15^\circ$) to avoid the Galactic disk, and those within $6^\circ$ from the Sagittarius-A20 \citep{Antoja2020,Ramos2020} sky track when the distance track is also between the limits along the line of sight defined for the target stream.

\subsection{Bayesian model} \label{sc:bayes}

In this study, we take the tracks reported in the literature as `true' and use them to find the RRL's membership probability of belonging to each stream. To do this, we constructed a Bayesian probabilistic mixture model in proper motions and distance, $\vec{y_n}=(\mu_{\phi_1},\mu_{\phi_2}, d)$. The details of the model are in Appendix \ref{ap:bayes}.

\subsection{Membership probability}

Once the posterior probability is sampled, we compute the membership probability for each star $n$, following \citet{Foreman-Mackey2014}, as the average in the posterior samples ($m$) of the probability of belonging to the stream over the total probability:
\begin{equation} \label{ec:pmemb}
    P_{\text{memb},n} \simeq \frac{1}{M}\sum\limits_{m=1}^{M} \frac{f^m p_{\text{st},n}^m}{f^m p_{\text{st},n}^m + (1-f^m)p_{\text{bkg},n}},
\end{equation}

\noindent where the weight $f$ is the stream versus background RRL ratio in the on-stream region (i.e., where the inference is made) and $M$ is the number of posterior samples. 
The complete list of RRL members for every track studied is provided in Table \ref{tb:membs}.

\subsubsection{Contamination}\label{sc:contamination}

The mixture model considers the existence of a background and models it in proper motion and distance (see Appendix~\ref{ap:bayes}). However, due to possible background variations along the stream, we also consider possible contamination from RRL that may have consistent proper motion and distance with the stream but are not real stream members. To estimate this, we count the number of RRL with a membership probability higher than 50\% outside the extended stream region in the sky. Assuming that proper motion and distance are locally independent of their sky position, we scaled this number by the sky area to determine the expected number of contaminants or false positive members according to our model, within the stream region. With this procedure we cannot say which of the stars that we find with our method are truly members of the stream or just background stars that, by chance, align with it; however, this provides a statistical estimate of the number of background RRL that we expect to have wrongly selected as members.

\section{Results and discussion} \label{sc:result}

\subsection{Summary of results} \label{sc:result_summary}

The analysis was performed by applying the method described in Sec.~\ref{sc:method1} to the 75 tracks in \verb|galstreams| with proper motion and distance information (corresponding to 56 streams, 12 of which have more than one track). In what follows, we present tables summarizing the results of the census for the full set of streams and for each stream track.

Table~\ref{tb:resumen_st} summarizes the numbers of streams with a certain amount of RRL in their tails, differentiating between those with or without progenitor. Table~\ref{tb:resumen_results} presents, for each stream track, the number of RRab and RRc identified in the progenitor (when it exists) and in the tails, the number of contaminants estimated as described in Sec.~\ref{sc:contamination}, and the inferred $f$ and intrinsic dispersions in proper motion and distance. Uncertainties for these quantities were computed as the standard deviation of the marginalized posterior probability of each quantity. 
Five tracks presented too few RRL in their background for the method to be applied reliably (C-4-I21, C-5-I21, Gaia-2-I21, M5-I21, Phoenix-S19) and were not taken into account for the analysis.

In what follows, we focus the discussion on the streams with~$\geq3$~RRL member identified and separate the discussion into two cases: streams with an associated progenitor and streams without one. In each case, we discuss according to the number of RRL detected in the tails. For each stream, we analyze the track adopted by default in \verb|galstreams|, unless specified otherwise.

\begin{table}
    \caption{Summary of the results: number of streams classified according to how many RRL we have detected in them.} \label{tb:resumen_st}
    
    \centerline{
    \small
    \resizebox{0.5\textwidth}{!}{%
    \begin{tabular}{c|ccccccccc}
         \hline\hline
         & $>$3 RRL & 1-3 RRL & RRL in & No RRL  & Not \\
         &in tails & in tails & prog. only & & reliable  \\
         \hline
         With progenitor &5&8&3&1&0 \\
         Without progenitor &9&11&-&15&4 \\
         
         \hline\hline
    \end{tabular}
    }
    }
\end{table}

\begin{table*}[h!]
    \caption{Summary of results for each track (with enough RRL in their background to be able to apply the method reliably). We show all the tracks with 3 o more RRL members ($RR_{ab}+RR_c$) in the stream's tails in descending order per track, the complete table is available in the online version. (1) Track name on \texttt{galstreams} (2) Corresponding stream name (3) Number of RRab in the tails (\textit{t}) and the progenitor (\textit{p}) (4) Number of RRc in the tails (\textit{t}) and the progenitor (\textit{p}) (5) Number of expected contaminants in the tails (6) Stream vs. background RRL ratio in the on-stream region (7) Distance intrinsic dispersion (8) $\phi_1$-proper motion intrinsic dispersion (9) $\phi_2$-proper motion intrinsic dispersion (10)~Reference of the track.} \label{tb:resumen_results}

    \centerline{
    \resizebox{1.05\textwidth}{!}{%
    \begin{tabular}{ccccccccccccc}
         \hline\hline
         TrackName & StreamName & $RR_{ab}$ & $RR_c$& $N_{\text{cont}}$& $d_\text{mean}$ & $f$ & $\sigma^{int}_d$ & $\sigma^{int}_{\mu_{\phi_1}}$ & $\sigma^{int}_{\mu_{\phi_2}}$ & Ref.\\
        &&\textit{t, p}&\textit{t, p}&&(kpc)&&(kpc)&(mas/yr)&(mas/yr)& (*) \\
        \hline\\
        Orphan-K19 & Orphan-Chenab & 103, - & 29, - & 12.19 & 24.14 $\pm$ 0.12 & 0.0698 $\pm$ 0.0065 & 2.81 $\pm$ 0.18 & 0.261 $\pm$ 0.019 & 0.138 $\pm$ 0.015 & (8) \\ 
        Orphan-K23 & Orphan-Chenab & 86, - & 29, - & 9.25 & 24.11 $\pm$ 0.12 & 0.0792 $\pm$ 0.0081 & 1.91 $\pm$ 0.29 & 0.262 $\pm$ 0.024 & 0.104 $\pm$ 0.019 & (9) \\ 
        Orphan-I21 & Orphan-Chenab & 24, - & 10, - & 1.3 & 17.99 $\pm$ 0.16 & 0.071 $\pm$ 0.013 & 2.26 $\pm$ 0.12 & 0.237 $\pm$ 0.013 & 0.288 $\pm$ 0.015 & (7) \\ 
        Chenab-S19 & Orphan-Chenab & 2, - & 0, - & 1.22 & 40.1 $\pm$ 1.4 & 0.23 $\pm$ 0.13 & 2.56 $\pm$ 1.98 & 0.035 $\pm$ 0.020 & 0.039 $\pm$ 0.021 & (16) \\ 
        Cetus-Palca-T21 & Cetus-Palca & 52, - & 25, - & 12.67 & 30.17 $\pm$ 0.20 & 0.0310 $\pm$ 0.0060 & 1.60 $\pm$ 0.13 & 0.038 $\pm$ 0.012 & 0.148 $\pm$ 0.010 & (17) \\ 
        Cetus-New-Y21 & Cetus-New & 2, - & 2, - & 0.0 & 16.60 $\pm$ 0.52 & 0.0030 $\pm$ 0.0020 & 4.30 $\pm$ 2.51 & 0.31 $\pm$ 0.19 & 0.17 $\pm$ 0.11 & (20) \\ 
        LMS1-M21 & Wukong & 13, - & 11, - & 6.03 & 17.48 $\pm$ 0.18 & 0.066 $\pm$ 0.016 & 2.88 $\pm$ 0.72 & 0.175 $\pm$ 0.046 & 0.183 $\pm$ 0.045 & (11) \\ 
        Pal5-PW19 & Pal 5 & 15, 1 & 5, 5 & 1.67 & 19.56 $\pm$ 0.21 & 0.131 $\pm$ 0.027 & 1.06 $\pm$ 0.52 & 0.028 $\pm$ 0.022 & 0.032 $\pm$ 0.023 & (15) \\ 
        Pal5-I21 & Pal 5 & 15, 1 & 5, 5 & 0.51 & 18.86 $\pm$ 0.20 & 0.237 $\pm$ 0.047 & 2.41 $\pm$ 0.37 & 0.108 $\pm$ 0.044 & 0.031 $\pm$ 0.025 & (7) \\ 
        Turranburra-S19 & Turranburra & 13, - & 1, - & 1.76 & 26.19 $\pm$ 0.35 & 0.59 $\pm$ 0.12 & 2.71 $\pm$ 0.49 & 0.151 $\pm$ 0.048 & 0.176 $\pm$ 0.028 & (16) \\ 
        M92-I21 & M92 & 8, 7 & 3, 1 & 5.36 & 8.00 $\pm$ 0.10 & 0.073 $\pm$ 0.022 & 1.01 $\pm$ 0.15 & 0.798 $\pm$ 0.064 & 0.755 $\pm$ 0.059 & (7) \\ 
        AAU-ATLAS-L21 & ATLAS-Aliqa Uma & 8, - & 2, - & 0.0 & 19.40 $\pm$ 0.33 & 0.319 $\pm$ 0.085 & 1.76 $\pm$ 0.20 & 0.056 $\pm$ 0.015 & 0.068 $\pm$ 0.012 & (10) \\ 
        AAU-AliqaUma-L21 & ATLAS-Aliqa Uma & 0, - & 1, - & 0.45 & 25.9 $\pm$ 1.9 & 0.221 $\pm$ 0.14 & 1.66 $\pm$ 0.31 & 0.095 $\pm$ 0.023 & 0.097 $\pm$ 0.025 & (10) \\ 
        ATLAS-I21 & ATLAS-Aliqa Uma & 7, - & 2, - & 0.32 & 20.29 $\pm$ 0.36 & 0.266 $\pm$ 0.077 & 4.07 $\pm$ 0.48 & 0.205 $\pm$ 0.028 & 0.162 $\pm$ 0.026 & (7) \\ 
        Elqui-S19 & Elqui & 8, - & 2, - & 0.75 & 48.28 $\pm$ 0.91 & 0.71 $\pm$ 0.18 & 4.18 $\pm$ 0.58 & 0.100 $\pm$ 0.037 & 0.104 $\pm$ 0.034 & (16) \\ 
        Jhelum-b-B19 & Jhelum & 4, - & 2, - & 0.0 & 11.77 $\pm$ 0.25 & 0.071 $\pm$ 0.031 & 1.51 $\pm$ 0.98 & 0.38 $\pm$ 0.14 & 0.39 $\pm$ 0.14 & (1) \\ 
        Jhelum-a-B19 & Jhelum & 3, - & 0, - & 1.09 & 11.57 $\pm$ 0.34 & 0.069 $\pm$ 0.054 & 2.4 $\pm$ 2.0 & 0.45 $\pm$ 0.22 & 0.38 $\pm$ 0.20 & (1) \\ 
        Jhelum-I21 & Jhelum & 3, - & 2, - & 1.05 & 11.40 $\pm$ 0.27 & 0.111 $\pm$ 0.050 & 0.35 $\pm$ 0.13 & 0.552 $\pm$ 0.038 & 0.489 $\pm$ 0.035 & (7) \\ 
        Jhelum-b-S19 & Jhelum & 2, - & 1, - & 0.0 & 10.86 $\pm$ 0.32 & 0.192 $\pm$ 0.092 & 2.20 $\pm$ 0.66 & 0.171 $\pm$ 0.039 & 0.150 $\pm$ 0.035 & (16) \\ 
        Jhelum-a-S19 & Jhelum & 2, - & 1, - & 0.0 & 12.49 $\pm$ 0.36 & 0.061 $\pm$ 0.033 & 1.00 $\pm$ 0.41 & 0.134 $\pm$ 0.027 & 0.190 $\pm$ 0.041 & (16) \\ 
        OmegaCen-I21 & Omega Centauri & 4, 32 & 2, 45 & 3.6 & 4.80 $\pm$ 0.03 & 0.183 $\pm$ 0.020 & 0.903 $\pm$ 0.088 & 0.500 $\pm$ 0.045 & 0.421 $\pm$ 0.046 & (7) \\ 
        Fimbulthul-I21 & Omega Centauri & 2, 34 & 1, 45 & 2.32 & 4.86 $\pm$ 0.03 & 0.240 $\pm$ 0.025 & 0.497 $\pm$ 0.058 & 1.69 $\pm$ 0.13 & 0.504 $\pm$ 0.047 & (7) \\ 
        TucanaIII-S19 & Tucana III & 4, 0 & 2, 0 & 0.78 & 20.12 $\pm$ 0.39 & 0.37 $\pm$ 0.18 & 5.2 $\pm$ 1.9 & 0.067 $\pm$ 0.029 & 0.068 $\pm$ 0.028 & (16) \\ 
        NGC1851-I21 & NGC 1851 & 4, 8 & 1, 5 & 0.67 & 11.99 $\pm$ 0.15 & 0.106 $\pm$ 0.027 & 0.21 $\pm$ 0.25 & 0.141 $\pm$ 0.044 & 0.153 $\pm$ 0.044 & (7) \\ 
        GD-1-I21 & GD-1 & 5, - & 0, - & 1.69 & 8.69 $\pm$ 0.19 & 0.0151 $\pm$ 0.0083 & 0.610 $\pm$ 0.037 & 0.450 $\pm$ 0.012 & 0.338 $\pm$ 0.010 & (7) \\ 
        GD-1-PB18 & GD-1 & 3, - & 0, - & 0.56 & 8.10 $\pm$ 0.20 & 0.021 $\pm$ 0.011 & 1.35 $\pm$ 1.44 & 0.33 $\pm$ 0.17 & 0.21 $\pm$ 0.14 & (14) \\ 
        Indus-S19 & Indus & 3, - & 1, - & 0.36 & 14.98 $\pm$ 0.35 & 0.022 $\pm$ 0.020 & 4.10 $\pm$ 0.49 & 0.145 $\pm$ 0.030 & 0.109 $\pm$ 0.020 & (16) \\ 
        M3-Y23 & M3-Svöl & 4, 97 & 0, 19 & 0.69 & 9.75 $\pm$ 0.04 & 0.531 $\pm$ 0.032 & 1.08 $\pm$ 0.08 & 0.192 $\pm$ 0.013 & 0.182 $\pm$ 0.012 & (19) \\ 
        Svol-I21 & M3-Svöl & 0, - & 0, - & 0.0 & - & 0.0050 $\pm$ 0.0070 & 0.331 $\pm$ 0.088 & 0.777 $\pm$ 0.052 & 0.584 $\pm$ 0.041 & (7) \\ 
        Phlegethon-I21 & Phlegethon & 2, - & 1, - & 0.0 & 3.45 $\pm$ 0.10 & 0.0140 $\pm$ 0.0091 & 0.5 $\pm$ 1.2 & 1.136 $\pm$ 0.042 & 0.395 $\pm$ 0.018 & (7) \\ 
        NGC3201-P21 & NGC 3201-Gjöll & 3, 68 & 0, 5 & 1.84 & 4.40 $\pm$ 0.02 & 0.414 $\pm$ 0.037 & 0.84 $\pm$ 0.06 & 0.996 $\pm$ 0.067 & 0.279 $\pm$ 0.016 & (13) \\ 
        Gjoll-I21 & NGC 3201-Gjöll & 2, 0 & 0, 0 & 0.0 & 3.51 $\pm$ 0.13 & 0.043 $\pm$ 0.035 & 0.209 $\pm$ 0.041 & 1.013 $\pm$ 0.075 & 0.261 $\pm$ 0.023 & (7) \\ 
        NGC3201-I21 & NGC 3201-Gjöll & 0, 68 & 0, 5 & 0.34 & 4.42 $\pm$ 0.02 & 0.782 $\pm$ 0.043 & 0.59 $\pm$ 0.04 & 1.084 $\pm$ 0.043 & 0.510 $\pm$ 0.022 & (7) \\ 

        \multicolumn{11}{c}{\dots} \\
        \hline
        \hline
        \end{tabular}
        }}
        \vspace{0.25cm}
{\small (*) \textbf{References:} (1): \citet{Bonaca2019}, (2): \citet{Caldwell2020}, (3): \citet{Chandra2022}, (4): \citet{Ferguson2022}, (5):~\citet{Grillmair2019}, (6): \citet{Grillmair2022}, (7): \citetalias{Ibata2021}, (8): \citet{Koposov2019}, (9): \citet{Koposov2023}, (10): \citet{Li2021}, (11): \citet{Malhan2021b}, (12): \citet{Palau2019}, (13): \cite{Palau2021},  (14): \citet{Price-Whelan2018}, (15): \citet{Price-Whelan2019}, (16):~\citetalias{Shipp2019}, (17): \citet{Thomas2022}, (18): \citet{Williams2011}, (19): \citet{Yang2023}, (20):~\citet{Yuan2022}.}
\end{table*}

\subsection{Streams with an associated progenitor}
We first discuss the 17 streams (25 tracks) with an associated progenitor. Out of these, four streams did not have any RRL in their tails (NGC 288, NGC 2298, NGC 2808 and NGC 6397), and in particular, NGC 6397 did not have any RRL at all.

In order to discuss RRL within the cluster or in the tails of the stream, we adopted the tidal radius described by Eq. 8 of \citet{Webb2013}, which takes into account the orbit and the orbital phase of the cluster. When compared with previous studies, some adopted different criteria (e.g., the value of the tidal radius at the pericenter). This means that there will be some differences in the number of RRL in and out of the cluster. However, unless the differences have a significant impact on the interpretation, they are not discussed.

\subsubsection{Tucana III} \label{sc:tucanaiii}

Tucana III is an ultra-faint dwarf galaxy\footnote{Although there is still uncertainty about its nature \citep{Li2018}.} (UFD) and, in our sample, the only stream with a surviving associated progenitor that is not a globular cluster. This UFD has a luminosity of $L_{\mathrm{UFD}} \simeq 800L_\odot$, a stellar mass of $M_{\mathrm{UFD}} \simeq 0.8\times 10^3M_\odot$ \citep{Drlica-Wagner2015} and a spectroscopic metallicity of [Fe/H]$=-2.42$~dex \citep{Simon2017}. The tracks used to study the stream are from \citet{Shipp2018}, who estimated the initial stellar mass of the progenitor to be $M_{i\mathrm{,prog}}=3.8\times 10^3M_\odot$ based on the observed stars and an assumed initial mass function. 

We found 6 RRL in the Tucana III stream (with an expected number of contaminants of 0.8) and none of them in the UFD itself, which is consistent with its low stellar mass. Three of the RRL have photometric metallicities measured by \citet{Li2023}, with values of –1.96, –2.30, and –2.40~dex (uncertainties $\sim0.3$~dex), all in agreement with the spectroscopic metallicity of the UFD. The stellar mass of the stream can be estimated from the number of RRL as follows: 
combining the number of RRL in globular clusters from \citet{Reyes2024} with their stellar masses from \citet{Baumgardt&Hilker2018} we get a production rate of one RRL per $1.39^{+4.25}_{-1.05}\times 10^4 M_\odot$; this implies that the stream’s progenitor lost on the order of $2$–$34 \times 10^4M_\odot$\footnote{This value is a lower limit, since we cannot rule out that additional RRL escaped earlier from the UFD and are now already phase-mixed.}, which would be nearly all of its stellar mass.

This value is significantly larger than the estimate of \citet{Shipp2018}. However, their system mass estimate is based on counting stars along the observed length of the stream and therefore provides only a lower limit to the progenitor mass. Portions of the tidal tails may lie outside the DES footprint, be too faint to detect, or have already become phase-mixed. It is therefore not surprising that their estimate may be underestimated, especially since we identify RRL extending farther along the stream track.

It is worth noting that if we use the mass-metallicity relation from \citet{Kirby2013} and the spectroscopic metallicity of the UFD, we obtain a mass of $	M_{\mathrm{prog}}=3.72 ^{+10.09}_{-2.72} \times 10^3~M_\odot$, consistent with \citet{Shipp2018} estimate. However, this mass-metallicity relation has a large dispersion. For example, two dwarf galaxies in the Local Group, Leo IV and Andromeda X, have nearly the same metallicity as Tucana III ([Fe/H]$=-2.45$~dex and [Fe/H]$=-2.46$~dex, respectively); yet, their stellar masses are significantly higher ($M_* \sim10^4M_\odot$ and $M_* \sim10^5M_\odot$, respectively). These values are consistent with our estimation for the Tucana III system.

In addition, \citet{Riley2025}, using the suite of Milky Way-mass halos from the Auriga project \citep{Grand2017,Grand2024}, found that  tidal disruption can shift satellites away from the intrinsic mass-metallicity relation. The mass-metallicity relation that is typically measured observationally corresponds to a tidally evolved relation, which differs from the intrinsic one because satellites lose stellar mass and preferentially lose low-metallicity stars from their outskirts. 
As a result, stellar masses inferred from the observed mass-metallicity relation may underestimate the true stellar mass of disrupting systems. Our lower bound for the mass of Tucana III implies an underestimation of $\log_{10}(3.72\times10^3/2.00\times10^4)=-0.73$, suggesting that the system has experienced substantial tidal mass loss, broadly consistent with this picture.

\begin{figure*}[h!]
    \centering
    \includegraphics[width=.8\linewidth]{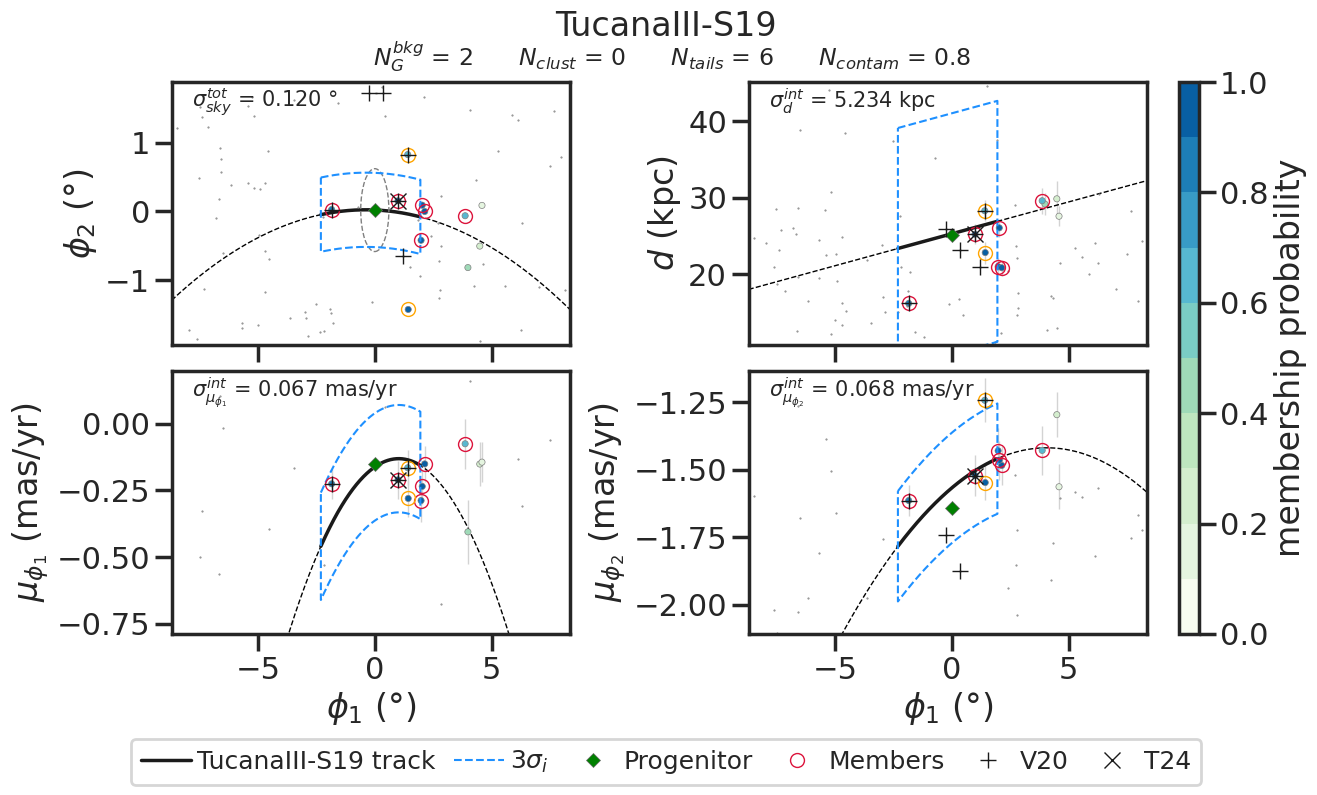}
    \caption{RRL stars in the vicinity of the Tucana III stream colored by membership probability with its tracks (solid black) and extrapolation (dashed black), RRL with membership probability $<$ 0.1 are plotted with a small gray marker. \textbf{Upper left:} Sky position (in the stream coordinate frame) of the RRL. \textbf{Upper right:} Distances as a function of $\phi_1$. \textbf{Lower panels:} Proper motions in $\phi_1$ and $\phi_2$ as a function of $\phi_1$. The blue dashed regions shown $3\sigma$ to each side of the track, with $\sigma$ equal to the observed width for the sky and to the intrinsic dispersions for the rest. In all panels, RRL with a membership probability higher than 50\% and closer than $3\sigma_{sky}^{tot}$ in the sky are marked as red circles and are considered members of the stream, while if they are further than $3\sigma_{sky}^{tot}$ are marked as orange circles and are used to estimate the number of contaminants within the members, the black plus symbols are RRL detected by \citet{Vivas2020} and the black cross is the RRL detected by \citet{Tau2024}. Additionally, we also show the number of Gaussians used for the background model, the number of RRL detected in the cluster and tails, and the number of expected contaminants.} \label{fig:TucIII_results}
\end{figure*}

To our knowledge, only two studies have searched for RRL around Tucana III (among other ultra-faint dwarf galaxies): \citet{Vivas2020} and \citet{Tau2024}. \citet{Vivas2020} used the Gaia~DR2 catalog of RRL \citep{Clementini2019} to select all stars within a 2$^\circ$ radius centered on Tucana III and compared their proper motions with those of spectroscopically confirmed members of the UFD.  
Using this approach, they found no RRL within Tucana III, but identified 6 extra-tidal candidates, shown in Fig. \ref{fig:TucIII_results}. However, as noted in their study, these stars do not appear to follow the tails of the stream but are instead distributed around the UFD. \citet{Tau2024} searched for extended stellar populations in UFDs following \citet{Vivas2020}, but they enhanced it by using the RRL catalogs from Gaia~DR3, Zwicky Transient Facility \citep[ZTF;][]{Huan2022}, Dark Energy Survey \citep[DES;][]{Stringer2021}, and PS1. 
They identified one RRL on the outskirts of Tucana III -- one of the same stars previously reported by \citet{Vivas2020} -- shown in Fig. \ref{fig:TucIII_results}.

Both studies based their methods on the argument that the population of RRL in the Milky Way declines with Galactocentric radius, becoming very rare beyond $\sim$50~kpc \citep[e.g.][]{Zinn2014, Hernitschek2017, Medina2018, Stringer2021}. Since most UFD lie at greater distances, contamination from the halo field stars can generally be neglected. However, Tucana III is located at a Galactocentric distance of only $\sim$23~kpc; therefore, this argument becomes less strong and a more rigorous method, like that used in this work, would provide a more robust association to the dwarf and the stellar stream. 

We recovered 3 of the RRL identified by \citet{Vivas2020} with a membership probability greater than 50\%. Of these, two are classified as members -- one of which is also the one reported by \citet{Tau2024} -- while the third is not, as it lies significantly far from the celestial track ($7.2\sigma_{\text{sky}}^{\text{tot}}$). Our model allows us to assess the reliability of detected members through their membership probabilities, which exceed 90\% for the two confirmed RRL in common. In contrast, the remaining three RRL candidates from \citet{Vivas2020} have membership probabilities lower than 10\%. As shown in Fig.~\ref{fig:TucIII_results}, not only are these stars spatially distant from the track on the sky, but they also have proper motions in $\phi_1$ that fall outside the range displayed in the figure. This is not surprising, as \citet{Vivas2020} had already noted that their candidates do not appear to follow the stream.

In summary, we have identified 6 RRL: 4 new detections and 2 previously reported as extra-tidal by \citet{Vivas2020} and \citet{Tau2024}. Of the 4 new detections, 2 have photometric metallicity, both consistent with the spectroscopic metallicity of the UFD, as mentioned above, making their classification as members even more trustworthy. In addition, all are in the extrapolation of the trailing arm of the stream, extending its length up to $\sim2^\circ$. Finally, of the remaining 4 RRL previously identified by \citet{Vivas2020}, 3 have membership probabilities $p_{\text{memb}} < 0.5$ and are therefore rejected as members by our model. The fourth has a $p_{\text{memb}} > 0.5$, but lies at $7.2\sigma_{\text{sky}}^{\text{tot}}$ from the celestial track, exceeding our sky selection box and is therefore, also not reported here as a member.

\subsubsection{M92} \label{sc:M92}

\begin{figure*}
    \centering
    \includegraphics[width=.8\linewidth]{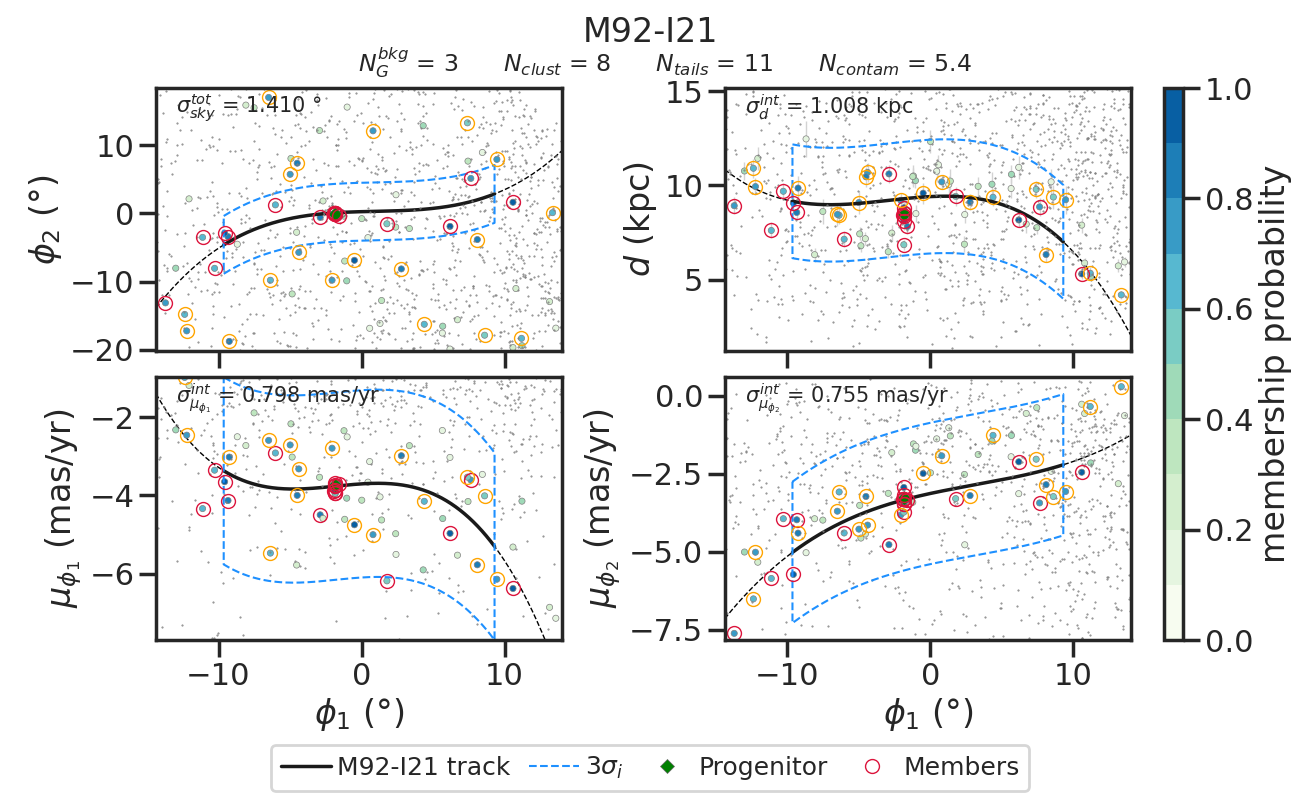}
    \caption{Similar to Fig. \ref{fig:TucIII_results} but for the M92 stream.} \label{fig:M92_results}
\end{figure*}

M92 (NGC 6341) is a metal-poor globular cluster \citep[$\text{[Fe/H] =}-2.35$~dex, ][]{Carretta2009} located at a heliocentric distance of 8.5~kpc and is currently near its apocenter \citep[$R_{\text{gal}}=9.85$~kpc, $R_{\text{peri}}=$1.25~kpc, $R_{\text{apo}}=$10.62~kpc, ][]{Baumgardt&Vasiliev2021}. Several tracks have been reported in the literature for the M92 stream 
(\citealt{Sollima2020}; \citealt{Thomas2020}, \citetalias{Thomas2020} hereafter; \citetalias{Ibata2021}; \citealt{Ibata2024}); however, only the track presented by \citetalias{Ibata2021} includes proper motion and distance measurements and is, therefore, the one adopted in this work. Figure~\ref{fig:M92_results} shows the results of the census using the \citetalias{Ibata2021} track. We identified 8 RRL within the cluster (compared to the 17 reported by \citealt{Reyes2024}, who did not impose RUWE quality constraints) and 11 candidates along the tidal tails of the M92 stream. We estimate that 5.4 of the latter are likely field contaminants.

\begin{figure*}[h!]
    \centering
    \includegraphics[width=.8\linewidth]{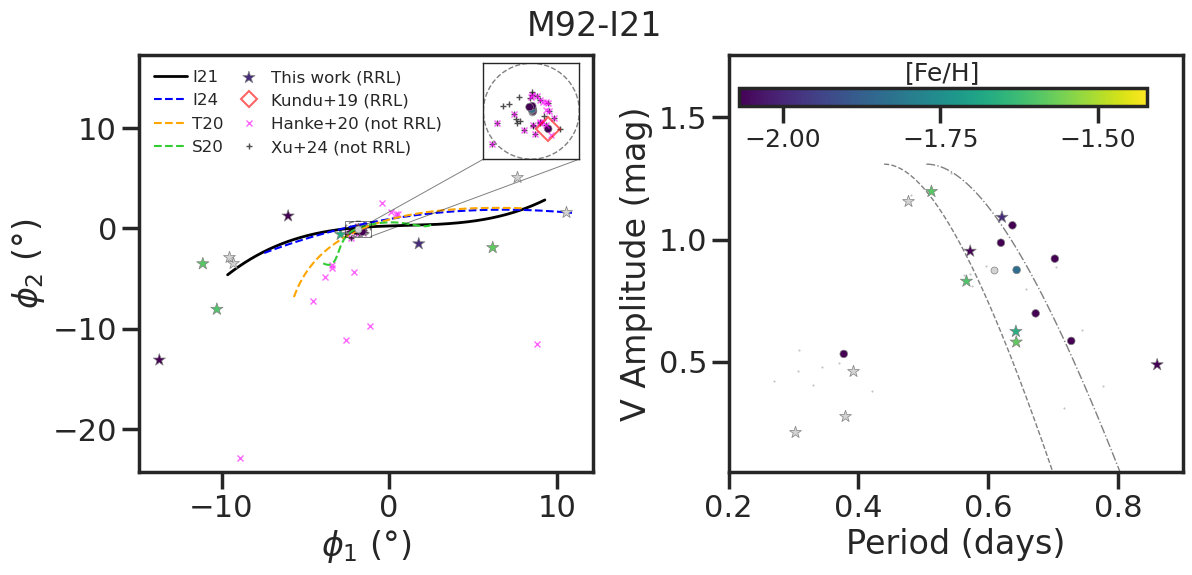}
    \caption{\textbf{Left:} Sky position (in the stream coordinate frame) of the vicinity of the M92 stream. The circles and star symbols are the RRL members identified by us colored by the photometric metallicity from \citet{Li2023}. Throughout both panels, star symbols denote RRL associated with the tidal tails and circles denote RRL associated to the globular cluster itself; both are colored by photometric metallicity from \citet{Li2023}. The red empty diamond is the RRL identified by \citet{Kundu2019} and recovered by \citet{Abbas2021}, the pink crosses and gray plus symbols are the members from \citet{Hanke2020} and \citet{Xu2024} respectively (in both cases these are not RRL). We also show the different tracks reported in the literature: green by \citet{Sollima2020}, orange by \citetalias{Thomas2020}, black by \citetalias{Ibata2021} (the one used for this work), and blue by \citet{Ibata2024}. \textbf{Right:} Amplitude vs. Period diagram of the RRL members that we identified in the vicinity of the M92 stream colored by its photometric metallicity. The period and amplitude uncertainties are $\sim5 \times 10^{-6}$ days and $\sim 0.02$ mag, respectively.} \label{fig:M92_AmpPFeH}
\end{figure*}

In previous studies, \citet{Kundu2019} and \citet{Abbas2021} identified 1 extra-tidal RRL nearby M92, which we recover as part of the cluster\footnote{They used a tidal radius of $r_t=30.5$ pc \citep{Moreno2014}. However, using a larger tidal radius of $r_t=113$~pc (which takes into account the eccentricity of the orbit and the orbital phase of the cluster, Eq. 8 of \citet{Webb2013}), that same star would be considered bound to the cluster.}, as shown in the left panel of Fig.~\ref{fig:M92_AmpPFeH}. \citet{Hanke2020} and \citet{Xu2024} searched for escapee stars; however, none of their candidates are RRL. In both cases, their candidates appear to follow the sky track presented by \citetalias{Thomas2020} and \citet{Sollima2020}, instead of the one from \citetalias{Ibata2021}, as shown in the left panel of Fig. \ref{fig:M92_AmpPFeH}.

In the right panel of Fig. \ref{fig:M92_AmpPFeH}, we show the Amplitude–Period diagram for the RRL that we have identified as members of the stream. A clear bimodality appears between the RRab stars within the cluster (circles) and those in the tails (star symbols), where the former show longer periods than the latter. This segregation could be explained by two possible factors: \textit{i)} at constant metallicity, more evolved RRL leave the zero-age horizontal branch, increasing their luminosity and therefore their period; \textit{ii)} at constant age, a metallicity gradient exists, where more metal-rich stars have shorter periods. In this case, the RRL in the tails lie in a higher metallicity locus of the diagram ($-1.6 \pm 0.1$~dex), consistent with their photometric metallicities from \citet[][determined from the $\phi_{31}$ Fourier parameter]{Li2023}; meanwhile, the RRL in the cluster lie in a more metal-poor locus ($-2.2 \pm 0.1$~dex), consistent with both their photometric metallicities and the spectroscopic metallicity of M92 \citep[$-2.35 \pm 0.05$~dex,][]{Carretta2009}. This coincidence, along with the fact that we expect all the stars in a globular cluster to have roughly the same age, supports the second explanation presented, and thus a clear segregation with a difference in metallicity of $\sim0.6$~dex between the tails and the cluster.

Since the metallicities of the RRL in the tails are similar to the mean metallicity of the halo background ($\sim-1.2$~dex), the question of whether these are genuine stream members or simply field contaminants (stars with proper motions and distances consistent with the stream but not physically associated with it) arises. To assess this possibility, we performed a control experiment using stars within the same box sample but located outside the on-stream and off-stream regions. This analysis yielded zero expected halo contaminants, suggesting that the 11 RRL identified in the stream region represent a significant overdensity with respect to the halo background.

On one hand, the metallicity dispersion obtained considering both the cluster and tail populations is too large ($\sim0.6$~dex) to have originated from a single globular cluster. On the other hand, the observed segregation implies a metallicity gradient that increases outward, opposite to what is typically seen in most of the dwarf galaxies in the Local Group \citep{Pilkington2012}. However, dwarf galaxies in the Centaurus group have been observed to have nuclei dominated by a Nuclear Star Cluster that is much more metal-poor than the rest of the galaxy and has a roughly simple star formation history \citep{Fahrion2022}. If this were the case for M92, it could represent the remnant Nuclear Star Cluster of a now-disrupted dwarf galaxy, with the stream tracing the \emph{galaxy's tidal debris} (similar to the case of the M54 globular cluster and the Sagittarius stream, although M54 presents a larger metallicity spread). In addition, the high metallicity of the tails ($\sim$-1.6~dex) implies a progenitor stellar mass of $\sim2\times10^6~M_\odot$, based on the mass-metallicity relation from \citet{Kirby2013}. 
This inferred mass places the progenitor within the dwarf galaxy regime rather than that of a globular cluster.

\citetalias{Thomas2020} proposed a similar scenario with a key difference: M92 was accreted with its host dwarf galaxy; however, the tidal debris was produced by the \emph{cluster itself}, and the host galaxy is now totally phase-mixed or in a completely different orbit. The authors proposed this origin history to explain the discrepancy between the dynamical age of the stream ($\sim 500$ Myr according to them) and the stellar age of the cluster \citep[between 11.0 and 13.80 Gyr, depending on the study,][]{Ying2023}. 
The two scenarios are not necessarily contradictory. Since the track proposed by \citetalias{Thomas2020} is not the same as that proposed by \citetalias{Ibata2021}, \citetalias{Thomas2020}'s track might be tracing the cluster's tidal tails, while \citetalias{Ibata2021}'s might trace the galaxy's tidal tails. 
Similar interpretations have been proposed for other stellar streams (e.g., Jhelum, \citealt{Bonaca2019}; LMS-1 with NGC 5024 and NGC 5053, \citealt{Yuan2020}; Cetus-Palca with Triangulum/Pisces, Turbio, Willka Yaku and C-20, \citealt{Yuan2022,Thomas2022}; GD-1, \citealt{Valluri2025}; C-19, \citealt{Yuan2025}).
Alternatively, simulations from \citetalias{Thomas2020} suggest that stars stripped during earlier pericenter passages ($\gtrsim 200$~Myr ago) present fanning and an offset from \citetalias{Thomas2020}'s track, which may also have experienced an increase in velocity dispersion due to M92 passage through the Galactic bulge; this may be what \citetalias{Ibata2021} is detecting. This region was not covered by the Canada-France Imagining Survey \citep[CFIS,][]{Ibata2017b} used in \citetalias{Thomas2020}'s analysis, which may explain their non-detection of this branch. Nonetheless, this alternative would not explain the difference in chemistry that we detected.

Additionally, the intrinsic proper motion dispersions result in $\sigma_{\mu_{\phi_1}}^{int} = 0.798$ mas/yr and $\sigma_{\mu_{\phi_2}}^{int} = 0.755$ mas/yr, which, at the distance of M92 (8.5~kpc), correspond to an intrinsic tangential velocity dispersion of $\sigma_{vt_{\phi_1}}^{int} = 32.13$ km/s and $\sigma_{vt_{\phi_2}}^{int} = 30.41$~km/s. These values are significantly larger than both the internal velocity dispersion of the cluster \citep[$\sim$6.5 km/s,][]{Baumgardt&Hilker2018} and the typical velocity dispersion of globular cluster streams \citep[usually less than 5 km/s,][]{Li2022}. 
To check this, we performed two tests: 
\textit{i)} Running our model while fixing the intrinsic velocity dispersion to match that of the M92 cluster; 
\textit{ii)} Running a more conservative prior based on the stream velocity dispersion distribution from \citet[][same as Eq. \ref{eq:priorpm} but with $\alpha=1.28427$ and $\beta= 0.15706$]{Li2022}, which allows high dispersions but favors smaller values. 
In both cases, no RRL were identified as stream members. We then repeated the experiments including the 84 confirmed stream members from \citetalias{Ibata2021} in our RRL sample. Under the fixed-dispersion model, we recovered less than 35\% of them. In contrast, when using the conservative prior, the intrinsic velocity dispersion converged to the one given by the \citetalias{Ibata2021} members ($\sim$30 km/s, consistent with our findings), and we recovered $\sim$90\% of their members as well as all the RRL that we had originally identified. These results indicate that the data from \citetalias{Ibata2021} require a large intrinsic velocity dispersion. Moreover, the low number of candidate RRL members prevents the inference from converging on this dispersion on its own, but this does not imply that these RRL are contaminants, as already addressed through our control experiments outside the on-stream and off-stream regions.

It is important to note that the intrinsic dispersion we measured is accurate only if the uncertainties of the observations are correctly estimated. There are a few ways this assumption could fail. On one hand, the proper motion errors might be underestimated. However, for this to really account for the observed dispersion, the errors would have to be significantly ($\sim35$ times) larger than the reported, an unlikely scenario given the precision of Gaia data. 
On the other hand, there could be a real gradient in distance and/or tangential velocity that is steeper than the one modeled on the stream tracks, which would translate as a large apparent dispersion in proper motion. To get a better handle on this, a more numerous tracer than RRL would be needed for which reasonably accurate distances could be measured. However, if the intrinsic dispersion obtained truly reflects the internal velocity dispersion of the stream, such a high value would imply a large mass, inconsistent with a globular cluster origin. This points to the same conclusions drawn from the metallicity measurements via the mass–metallicity relation, further supporting our scenario in which the stream we detect is the remnant of a disrupted dwarf galaxy that once hosted M92.
Further analysis is needed to confirm this origin scenario; however, it is beyond the scope of this paper and will be addressed in future work.

\subsubsection{NGC 3201-Gjöll}
NGC 3201 is a globular cluster with a mass of $1.93\times10^5~M_\odot$ \citep{Baumgardt&Hilker2018}, a metallicity of -1.51~dex \citep{Carretta2009}, and a total of 87 RRL according to \citet{Reyes2024}. 

We identified 3 RRL in the tails and 73 in the cluster. 
The difference in the number of cluster RRL with respect to the 87 identified by \citet{Reyes2024} arises from the 12 RRL with RUWE $>$ 1.4 that we do not count, and the two additional RRL that are not in our catalog (one detected by them and the other detected by \citealt{Clement2017}). 
We estimate the stellar mass of the cluster to be between $2.6\times10^5M_\odot$ and $4.9\times10^6M_\odot$; the lower end of this distribution is consistent with the mass reported by \citep{Baumgardt&Hilker2018}.

\citet{Kundu2019} already had searched for extra-tidal RRL nearby NGC 3201, identifying 13 of them. 
Using the tidal radius adopted in this work (82.34 pc) instead of the one they used \citep[36.13 pc,][]{Moreno2014}, we consider all these stars to still be within the cluster. However, they appear to be in a more elongated structure with respect to the rest of the RRL of the cluster, suggesting that they are going to be the next totally unbounded RRL from NGC 3201. 
Their photometric metallicities are also consistent with those of the cluster and with its spectroscopic metallicity. The only peculiarity is their asymmetrical spatial distribution, with 4 in the direction of the leading arm and 9 in the direction of the trailing arm. This asymmetry was also observed by \citet{Kundu2019}, who attributed it to the combined effects of tidal disruption with stripped debris from NGC 3201.

\citet{Hansen2020} chemo-dynamically linked the NGC 3201 cluster with the Gjöll stream, and identified 2 RRL as stream members in the process. These correspond to 2 of the 3 RRL that we identified as members of the tails. All 3 RRL are located at the end of the detected trailing arm. The absence of RRL closer to the cluster is not surprising because of the presence of the Galactic disk. However, given the elongated structure on the outskirts of the cluster, one might expect to find RRL in this region as well. With respect to the mass of the stream, using the lower end of the RRL production rate distribution to be consistent with the cluster, we estimate a minimum stellar mass of $1.02\times10^4M_\odot$ for the trailing arm.

In summary, we identify a new RRL member in the trailing arm of the stream, with a membership probability of $p_{\text{memb}}=0.992$. The identified RRL members do not show any segregation in type or metallicity, apart from the asymmetrical distribution in the elongated structure on the outskirts of NGC 3201.

\subsubsection{Palomar 5}
Pal 5 was the first stellar stream associated with a globular cluster to be discovered \citep{Odenkirchen2001, Rockosi2002}.

It is one of the few stellar streams that has been characterized using RRL \citep{Price-Whelan2019}. 
Using Gaia~DR2 data, they identified 10 RRL within the tidal radius of the cluster ($11^\prime$ according to them) and almost twice as many (17) located in the tails. Furthermore, they observed a segregation in the RRL population of the system: While most of the RRc (8 out of 12) are in the cluster, most of the RRab (13 out of 15) are in the tails.

Adopting a tidal radius of 49.73 pc (corresponding to $7.8^\prime$ at the distance adopted), we identified 6 RRL in the cluster and 20 in the tidal tails (expecting 1.7 of them to be contaminants) based on the track from \citet{Price-Whelan2019}. We recover the strong segregation by RRL type previously reported by \citet{Price-Whelan2019}: 5 out of the 6 RRL of the cluster are RRc, while 15 out of the 20 RRL of the tails are RRab. We only find this kind of segregation in Pal 5; therefore, it might be explained simply by low number statistics.

Of the cluster RRL, we were able to recover all except one of those identified by \citet{Price-Whelan2019}\footnote{The other 3 RRL that \citet{Price-Whelan2019} consider from the cluster are part of the tidal tails under our criteria.} -- and \citet{Reyes2024}. The missing star is the furthest from the cluster in terms of distance; however, this is due to the slightly richer metallicity assigned from the halo metallicity distribution (see Sec. \ref{sc:data}). Of the tails RRL, we recover most of the RRL identified by \citet{Price-Whelan2019}, except for 3 in the trailing arm, which, due to the updated value of their distances, are now less consistent with the track. Furthermore, we identified 3 new RRL: 2 located near the cluster in the trailing arm, and 1 in the extended track of the leading arm. The latter RRL is consistent with the track identified by \citet{Starkman2020}, providing new distance and proper motion information for that segment of the stream, which previously only had positional information.

\subsubsection{NGC 1851} \label{sc:NGC1851}

The tracks identified by \citetalias{Ibata2021} show an offset with respect to the cluster in distance and proper motion ($\Delta d=-1.8$~kpc, $\Delta \mu_{\phi_1}=-0.12$ mas/yr, and $\Delta \mu_{\phi_2}=0.18$ mas/yr; see Fig. \ref{fig:NGC1851_results})
. We therefore apply a manual correction to make them match with the cluster before applying our method. We identified 13 RRL in the cluster and 5 in the tails (expecting 0.7 of them to be contaminants); 2 out of the 5 tail members are new detections. Regarding the cluster members, we recovered all but one of the RRL with RUWE $<$ 1.4 \citep[28 in total,][]{Reyes2024}.

\citet{Abbas2021} previously searched for extra-tidal RRL in the vicinity of NGC 1851, visually classifying 5 stars as candidate members of the cluster: 2 with high confidence and 3 with intermediate confidence. 
We confirmed the 2 high-confidence RRL candidates located closer to the cluster, as well as 2 of the 3 intermediate-confidence ones. The one that we do not identify is due to its proper motion not being consistent with the stream. The other two also have a lower membership probability in our model ($p_{\text{memb}}=$ 0.654 and 0.584) compared to the rest of the identified members, which have $p_{\text{memb}}>0.88$. Furthermore, one of them (Gaia~DR3 4822412580148688896) has a photometric metallicity of [Fe/H]=$-2.53 \pm 0.30$~dex \citep{Li2023}, which is much more metal-poor than both the rest of the members ($\overline{\text{{[Fe/H]}}}=-1.59 \pm 0.17$~dex) and the spectroscopic metallicity of the NGC 1851 cluster \citep[$-1.18 \pm 0.08$~dex,][]{Carretta2009}, suggesting that this RRL might be a contaminant.

In order to assess whether correcting the tracks to match the cluster distance and proper motion was justified, we performed a control experiment. We first ran our method using the original (uncorrected) tracks on stars in the on-stream region (as usual), and second on stars within the same box sample (sky window plus distance limits) but located outside both the on-stream and off-stream regions (since this was used for the background model). In the first case, we identified 12 RRL in the tails; however, in the second case, we identified between 3 and 6, indicating a statistical overdensity of halo RRL consistent with the distance and proper motion of the uncorrected tracks, but which are not actual stream members. When we repeat the same experiment using the corrected tracks, the expected number of halo RRL is consistent with zero, confirming that our decision was appropriate. More importantly, this highlights that the method is very sensitive to the choice of tracks, and a reason to, in a next stage of this work, perform this analysis blindly (i.e., without assuming tracks) in the cases with enough RRL members \citep[as in the case of Pal 5,][]{Price-Whelan2019}.

\subsubsection{$\omega$Centauri-Fimbulthul}

 $\omega$Centauri ($\omega$Cen; NGC 5139) presents several peculiar features, including distinct multiple stellar populations with a wide range of ages and metallicities \citep{Johnson2010}, as well as different kinematics correlated with chemical abundances \citep{Bellini2018}.

It is important to note that while both tracks are connected to the cluster in proper motion and distance, they are not spatially connected to it. This suggests that the results using these tracks must be taken with caution, and highlights again the importance of performing this analysis blindly when the number of RRL members allows it.

\begin{figure*} 
    \centering
    \includegraphics[width=.8\linewidth]{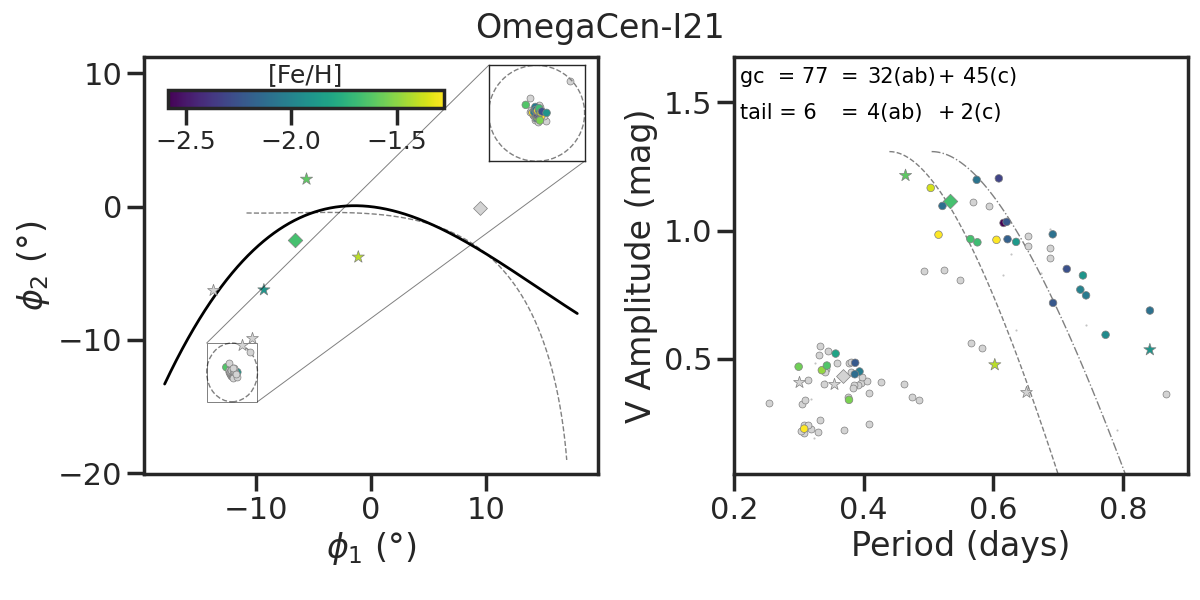}
    \caption{Similar to Fig. \ref{fig:M92_AmpPFeH} but for the $\omega$Cen stream. The diamond symbols correspond to the two additional RRL identified with the Fimbulthul track (gray dashed line of left panel) from \citet{Ibata2019}.}\label{fig:OmegaCen_AmpPFeH}
\end{figure*}

We recovered 77 out of the 84 RRL present in our catalog with RUWE$<$1.4 in the cluster \citep{Reyes2024} and identified 6 RRL in the tail debris, all of which are new detections; however, we expect 3.6 of them to be contaminants. Among the 77 RRL in the cluster, 29 have photometric metallicity available from \citet{Li2023}. We fitted a 1D Gaussian mixture model to their metallicity distribution using two components, according to the BIC fitting. 
The two populations are centered at -2.11 and -1.50~dex, the former being the dominant one. 
They correspond to the two groups of RRab visible in the right panel of Fig. \ref{fig:OmegaCen_AmpPFeH}. They are consistent with two of the four stellar populations identified by \citet{Kuzma2025} within a radial distance of $15^\prime$ from the cluster center, which show peaks at -2.11, -1.83, -1.50, and -1.19~dex (the middle two being the most dominant).  

Out of the 6 RRL identified in the tails, 3 have photometric metallicity: 2 are consistent with the more metal-rich population ([Fe/H]=$-1.60\pm 0.29$ and $-1.41 \pm 0.32$~dex), while the third ([Fe/H]=$-1.90 \pm 0.28$~dex) is consistent with the more metal-poor population; this behavior is also appreciated in the right panel of Fig. \ref{fig:OmegaCen_AmpPFeH}. We find no sign of spatial, type, or metallicity segregation within the cluster or the tails, although the number of RRL in the tails is too small to draw robust conclusions. Furthermore, using the track from \citet{Ibata2019}, we identified two additional RRL members in the tails (diamond symbols in Fig. \ref{fig:OmegaCen_AmpPFeH}); one of them has a photometric metallicity of [Fe/H]=$-1.66 \pm 0.67$~dex, which is also more consistent with the more metal-rich population.

In summary, within the cluster, the RRL stars are grouped into two populations with different metallicities ([Fe/H] = –2.11 and –1.50~dex). We identified 6 (+2) RRL stars in the tidal tails, whose photometric metallicities and positions in the period–amplitude diagram suggest they also belong to the same populations found within the cluster, with the more metal-rich population being dominant (opposite to what happens within the cluster). This is the first time RRL stars have been detected in the $\omega$Cen stream and associated with its distinct stellar populations. However, it is important to note that this analysis was made using tracks whose celestial components do not directly connect to the cluster, illustrating one more reason to perform this analysis blindly when possible.

\subsubsection{M3-Svöl}

We identified the 118 RRL with RUWE$<$1.4 within the cluster in our sample and only 4 RRL in the stream (expecting 0.7 of them to be contaminants); this is another case of a progenitor with a significant number of RRL within the cluster and just a few in the tidal tails. Previous studies have already identified RRL associated with M3. \citet{Kundu2019} identified 2 extra-tidal RRL (according to their classification), which we identified as part of the cluster. \citet{Hanke2020} identified one RRL; we found that this RRL has a $p_{\text{memb}}>0.5$ but it is located too far from the celestial track ($4.28\sigma_{\phi_2}$) to be considered a member. \citet{Wang2022} using integrals of motion to identify RRab associated with halo substructures found that 31 RRL are likely to be associated with M3; out of these, we identified only 3 with $p_{\text{memb}}>0.5$, including the one identified by \citet{Hanke2020}. The rest of the RRL are too far from the tracks on at least one of the components, and the majority of them follow the peak of the background proper motion distributions, supporting their rejection as stream members. One of these identified members (Gaia~DR3 3958056665299734144) has the highest photometric metallicity \citep[{[Fe/H]=$-0.73 \pm 0.37$~dex,}][]{Li2023} among the identified members, suggesting that this star is likely a contaminant.

\subsubsection{Remaining streams with an associated progenitor}
The other streams with an associated globular cluster studied in the present work are: M68-Fjörm, M2, M5, NGC 1261, NGC 5466 and NGC 6101. These streams have between 1 and 2 RRL in their tails each; therefore, there is not much to say about them on a statistical or population level.

\subsection{Streams without an associated progenitor} \label{sc:st_wo_prog}

Now we will discuss the 35 streams (45 tracks) without an associated progenitor. Out of these, 15 did not have any RRL detected: Aquarius, C7, C8, C19, Gaia7, Gaia8, Gaia9, Gaia11, Gunnthrá, Hrid, Kshir, Kwando, M68\footnote{This stream was identified by \citetalias{Ibata2021} and associated to the M68 cluster; however, due to discrepancies with the cluster and the other two tracks of the M68 stream in proper motion , it is not taken as a track of the tidal tails of M68, but as a new stream without a progenitor associated}, Slidr and Spectre.

\subsubsection{Orphan-Chenab}

The Orphan-Chenab stream was one of the first stellar streams discovered \citep{Grillmair2006c, Belokurov2007}. \citet{Koposov2019} has already performed a comprehensive analysis of the Orphan-Chenab's RRL population using Gaia~DR2 data; they identified 110 RRL of which 108 are in our catalog, all with RUWE$<$1.4 \citep[although 9 of these are members of the Sagittarius stream,][]{Ramos2021}. We focus our analysis in the tracks from \citet{Koposov2023}, who updated them from \citet{Koposov2019} using Gaia~EDR3 data.

We identified 118 RRL in the Orphan-Chenab stream: 75 of them are shared with \citet{Koposov2019}. Of the 33 that we did not recover, 6 are within the Sagittarius footprint (5 of them actually belonging to the Sagittarius stream \citep{Ramos2021} and 3 more within the Galactic disk portion that we removed to avoid contamination (see Sec. \ref{sc:selection}). The rest are located in the locus of the intersection of the tracks with the background in proper motion (15 in the left extreme and 9 in the right), which makes the separation of both components more difficult. Among our identified members, there are 8 that belong to the Sagittarius stream \citep{Ramos2021}, 4 of them detected by \citet{Koposov2019} also. Therefore, we identified 39 new detections that belong to the Orphan-Chenab stream.

\subsubsection{Jhelum} \label{sc:Jhelum}

Jhelum is a complex stellar stream with separated branches in sky, distance, and proper motion. We focus our analysis on the \citetalias{Shipp2019} track, which is the most consistent with the latest data \citep{Woudenberg2023, Viswanathan2023, Awad2023}. Component $a$ is consistent with the narrow component, and component $b$ is consistent with the broader one.

\begin{figure*}
    \centering
    \includegraphics[width=.8\linewidth]{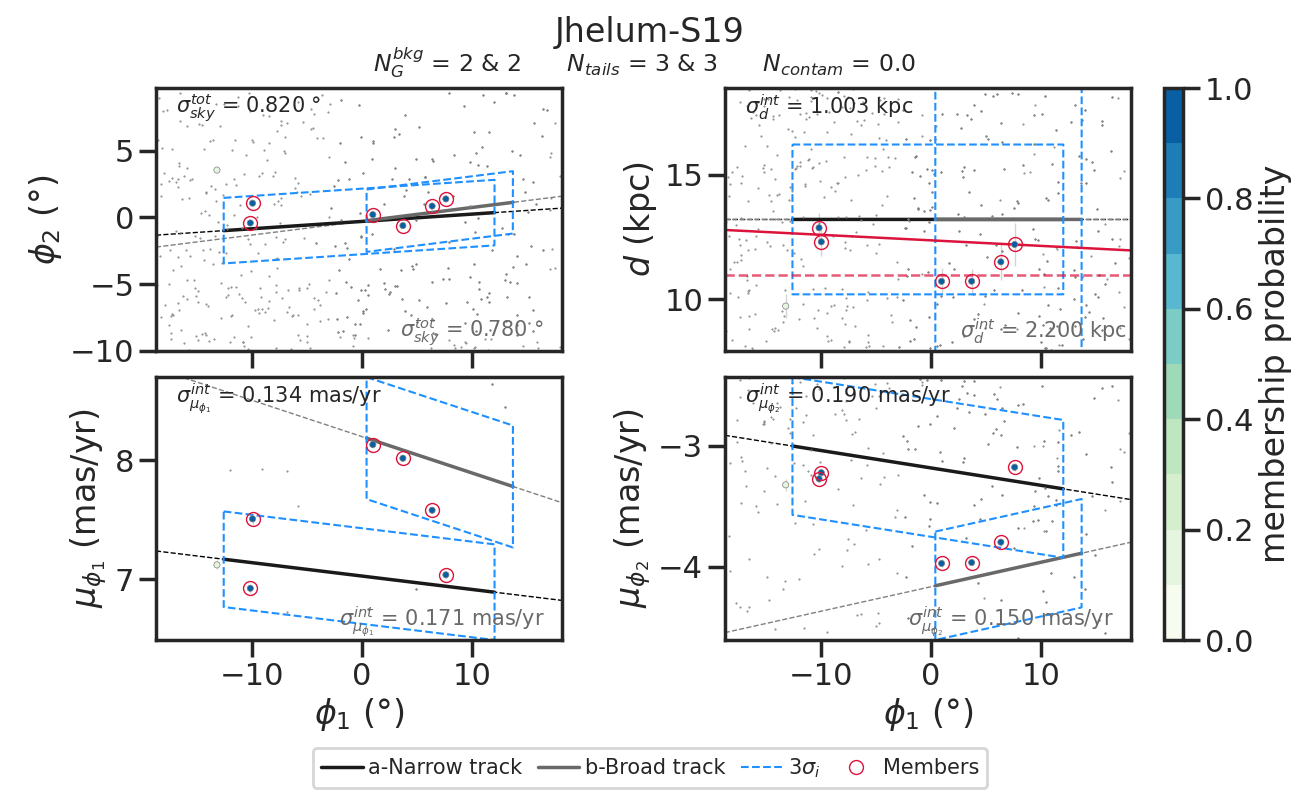}
    \caption{Similar to Fig. \ref{fig:TucIII_results} but for the Jhelum stream. The black track and dispersions correspond to the component $a$ of \citetalias{Shipp2019} (narrow component), and the gray ones to the component $b$ (broad component). The solid red line correspond to the inferred gradient distance for the narrow component, and the dashed light red line correspond to the mean distance of the broad component.}\label{fig:Jhelum_results}
\end{figure*}

We identified 3 RRL in the narrow component and 3 more in the broader one, all of which are new detections. The RRL of the broad component are too close together in $\phi_1$; therefore, we only infer a mean distance ($10.86 \pm 0.32$~kpc, dashed light red line in Fig.~\ref{fig:Jhelum_results}) instead of a distance track. This is different for the narrow component RRL, for which we fitted the 1D polynomial:
\begin{equation} \label{eq:dist_Jhelum}
    d \text{[kpc]} = 12.361 - 2.227\times10^{-2}\phi_1
\end{equation}
\noindent shown in Fig.~\ref{fig:Jhelum_results} as a solid red line, which has a mean distance of $d=12.49 \pm 0.36$~kpc. The mean values of the narrow and broad components are in perfect agreement with the mean values measured by \citet{Awad2023} for generic stars: 12.40~kpc for the narrow component and 10.95~kpc for the broad one. In addition, the mean distance of the narrow component is consistent with the mean values of \citet[][$d=13.00$~kpc]{Bonaca2019} and \citetalias[][$d=13.20$~kpc]{Shipp2019}, and the mean distance of the broad component is consistent with the mean value of \citet[][$d=10.41$~kpc]{Ibata2021}. We highlight the fact that we obtained two different values of distances for the two components, with the broader component located closer to the Sun than the narrow one; this is consistent with our method, even though the distance track used for both was the same.

The two components have different metallicities; we found that the narrow component is more metal-poor than the broad component ([Fe/H]=$-2.23 \pm 0.13$ and $-1.74 \pm 0.14$~dex, respectively), but with the same spread ($\sigma_{[Fe/H]}\simeq0.05$~dex). However, with only 3 stars per component, we cannot establish a robust value for the spread. These values are consistent with the spectroscopic members of Jhelum from \citet{Ji2020}, who, using $S^5$ data, observed 7 stars with a mean metallicity of -2.19~dex and a spread of 0.17~dex, as well as another star with -1.62~dex. On the other hand, we obtained a more metal-poor narrow component and a slightly more metal-rich broad component than \citet{Awad2023}, who, also using the metallicity data from the $S^5$ survey \citep{Li2019}, obtained a mean metallicity (and spread) of -1.87~dex (0.15~dex\footnote{This value is an upper limit of the metallicity spread for the narrow component.}) and -1.77~dex (0.34~dex) for the narrow and broad components, respectively. Although the mean values of \citet{Awad2023} differ from ours, both results support the scenario in which the Jhelum stream is the result of a globular cluster accreted along with its host dwarf galaxy, both of which are now disrupted, similar to the M92 case (see Sec. \ref{sc:M92}). In this scenario, the cluster and dwarf galaxy would be responsible, respectively, for the narrow and broad components. Nonetheless, alternative scenarios have been proposed in the literature to explain the morphology of Jhelum \citep[e.g.,][]{Bonaca2019, Woudenberg2023}; therefore, additional data are required to confirm this interpretation.

\subsubsection{Cetus-Palca / Cetus-New}

Cetus-Palca and Cetus-New are two different wraps of the same disrupted dwarf galaxy \citep{Chang2020,Yuan2022}. They occupy a similar portion of the sky, but Cetus-New is closer to the Sun than Cetus-Palca \citep[average distance of $d\sim18$~kpc and $d\sim40$~kpc respectively,][]{Yuan2022}.

We identified 77 RRL in the Cetus-Palca wrap, expecting 12.7 of them to be contaminants. There are 2 of them (Gaia~DR3 4618720144066419200 and 5221073742370884352) that belong to the Large Magellanic Cloud \citep{Soszynski2016} and another one (Gaia~DR3 2569266959934381696) that belongs to the Sagittarius stream \citep{Ramos2021}. The majority of the RRab members are Oosterhoff type II (OoII, 39 out of 52). The separation between both types is made using Eq.~2 of \citet{Belokurov2018}. The majority of the Oosterhoff type I (OoI, 9 out of 13) are located at the right extreme of the track, where the outskirts of the Magellanic Clouds are located; therefore, these stars are likely contaminants (one of these is one of those identified by \citealt{Soszynski2016}). The photometric metallicity of the RRL members is consistent with previous studies \citep{Yuan2022, Thomas2022}; we found a mean value of $-1.94 \pm 0.04$~dex and a spread of 0.44~dex.

In the Cetus-New wrap we identified 4 RRL, expecting no contaminants; however, the member Gaia~DR3 4618054183617627904 actually belongs to the Large Magellanic Cloud. The other 3 members consist of 1 RRab and 2 RRc, where only 2 of them have photometric metallicity, consistent with the metallicity of the Cetus-Palca warp.

\subsubsection{ATLAS-Aliqa Uma}

The ATLAS-Aliqa Uma (AAU) stream was proposed by \citet{Li2021} to be a single structure that unifies the previously identified ATLAS \citep{Koposov2014} and Aliqa Uma streams \citep{Shipp2018}. The stream appears discontinuous in the sky but remains coherent in distance, proper motion, and radial velocity.

We identified 10 RRL in the AAU stream, expecting 0.4 of them to be contaminants; 9 of these are in the ATLAS segment of the stream. \citet{Li2021} using $S^5$ spectroscopic data identified 5 RRL in the stream. We are able to recover all 4 RRL that lie on the ATLAS segment; the one that lies on the Aliqa Uma segment has a $p_{\text{memb}}>0.5$ but it is too far from the celestial track to be selected as a member. The authors also selected 6 more non-spectroscopic candidates that are consistent with the stream in position and distance (green circles on their Fig. 6), in particular, 2 RRL located at $\phi_1 \sim 6$. However, they did not publish which stars they are, and therefore, no further comparison can be made; although visually they match the location of two of our detections. In summary, we identified 10 RRL members, of which 6 are new detections.

The photometric metallicity of the RRL is consistent with the spectroscopic metallicity of the stream \citep[{[Fe/H]}=$-2.24$~dex,][]{Li2021}; they have a mean metallicity of [Fe/H]=$-2.21 \pm 0.10$~dex, with a spread of 0.12~dex. There is a subtle spread on the Oosterhoff type of the RRL: there are 8 RRab, of which 3 are type OoI and 5 are type OoII. Since the spread in the period-amplitude diagram is small (much smaller than in the M92 system), it could be attributed to the RRL leaving the zero-age horizontal branch, increasing their luminosities and, therefore, their periods (instead of a metallicity gradient); this also occurs in several globular clusters, of both Oosterhoff types, such as M5, M68 and M3 \citep[see Fig. 4 of][]{Cacciari2005}. There is no clear spatial segregation of the Oosterhoff type in the stream.

\subsubsection{GD-1}

We identified 5 RRL members, expecting 1.7 to be contaminants. There are two of these members (Gaia~DR3 1412046334799096704 and 1429088936827962752) that lie on the intersection of the stream tracks and the locus of the background in proper motion. In addition, their photometric metallicity ([Fe/H]=$-1.62\pm0.32$ and $-1.45\pm0.46$~dex) is very different from the metallicity of the stream \citep[{[Fe/H]} $\simeq$ -2.50~dex, ][]{Valluri2025}; therefore, they are likely contaminants. Although there is no study of RRL in GD-1 in particular, \citet{Tavangar2025} used Gaia~DR3 data to create a density map of the stream, where, within the sample they used, there are the other 3 RRL members that we identified successfully. Furthermore, if we use the track from \citet{Price-Whelan2018} instead of the one from \citetalias{Ibata2021}, we identify only these 3 RRL, supporting these as bona-fide members of the stream.

\subsubsection{LMS-1}

The "low-mass stream" (LMS-1) is a dwarf galaxy stellar stream \citep{Malhan2021}. 
Moreover, the globular clusters NGC~5024 and NGC 5053 are associated with the LMS-1 stream, where NGC 5054 is thought to be the nuclear star cluster of the dwarf galaxy progenitor of both the stream and the globular clusters \citep{Yuan2020}, similar to the scenario we propose for the M92 (Sec. \ref{sc:M92}) and Jhelum (Sec. \ref{sc:Jhelum}) streams.

\begin{figure}[h!]
    \centering
    \includegraphics[width=\linewidth]{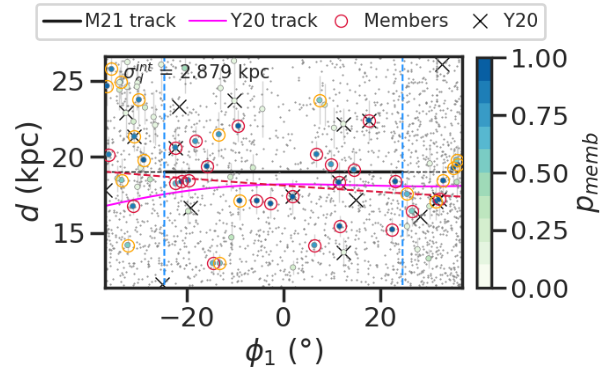}
    \caption{Distance as a function of $\phi_1$ for the LMS-1 stream. 
    The solid black lines are the mean distance reported by \citet{Malhan2021}. The magenta lines are the tracks from \citet{Yuan2020} and the black crosses are the RRL reported by them. The red dashed line are the distance track inferred from the RRL we identified (red circles), and the black dashed lines are the ones proposed by \citet{Li2022}.}\label{fig:LMS1_results}
\end{figure}

The default track in \verb|galstreams| is from \citet{Yuan2020}; however, this is too long to apply our method ($\sim175^\circ$); therefore, we analyzed the much shorter track from \citet[][$\sim50^\circ$]{Malhan2021}. Both are in excellent agreement \citep[fifth row of Fig. 5 from][]{Mateu2023}, although the track from \citet{Malhan2021} only has a mean distance (19.0~kpc). We identified 24 RRL of which 19 are new detections; however, we expect 6.0 of them to be contaminants.

\citet{Yuan2020} did the first report of LMS-1 identifying RRL and Blue Horizontal Branch stars (BHB) on the stream. They reported 20 RRL (black crosses in Fig. \ref{fig:LMS1_results}), of which we recovered only 5. It is important to mention that their RRL cover a projected width of $8.7^\circ$ and are likely to belong to multiple wraps of the stream \citep{Yuan2020}. The projected width of the \citet{Malhan2021} track is $1.7^\circ$; therefore, it is not surprising that we did not recover all the RRL from \citet{Yuan2020} since we focused on the analysis of only one wrap using the \citet{Malhan2021} track. \citet{Malhan2021} mention that they selected 91 RRL that follow the orbit of LMS-1; however, they did not publish a list of members; therefore, no further comparison is possible. 

With the identified RRL we fitted a distance track using a first degree polygonal\footnote{When a second degree polygonal is fitted, the quadratic coefficient is neglectable ($6.49\times10^{-5}$); therefore, we used a first degree polygonal.} (red dashed line of Fig.~\ref{fig:LMS1_results}):
\begin{equation}
    d \text{[kpc]} = 18.13 - 2.25\times10^{-2}\phi_1.
\end{equation}
\noindent The track is consistent with the mean value of \citet{Malhan2021}, and with the distance track of \citet{Yuan2020} in this segment of the stream ($-35^\circ < \phi_1 < 35^\circ$). However, since our inferred distance track has a negative slope, this behavior causes the track to differ from that of \citet{Yuan2020} for $\phi_1 \lesssim -15^\circ$. This stream is another example of a good candidate for a blind search.

\subsubsection{Turranburra} \label{sc:turranburra}

We identified 14 RRL in the stream, expecting 1.8 of them to be contaminants. Among the members, there is one (Gaia~DR3 5091015982252983680) that has a photometric metallicity of [Fe/H]=$-4.12 \pm 0.27$~dex \citep{Li2023}. Never before has such a metal-poor RRL been spectroscopically observed (only a few stars); therefore, this must be an extrapolation of the metallicity-$\phi_{31}$ relation in a regime in which it is probably not valid. 
The other members have a mean photometric metallicity of $-1.75\pm0.08$~dex with a spread of 0.12~dex -- higher than the one proposed by \citet[][{[Fe/H]}=$-2.18$ dex]{Li2022}, but in more agreement with the one estimated by \citet[][Z=0.0003, i.e. {[Fe/H]}=$-1.67$ dex using $Z_\odot=0.014$]{Shipp2018}. \citetalias{Shipp2019} selected 13 RRL around the Turranburra stream based on sky ($|\phi_1|<8.5^\circ$, $|\phi_2|<5^\circ$) and distance ($|\Delta d|<3$~kpc with respect to mean distance), shown as black crosses in the top panel of Fig.~\ref{fig:dist_gradients}; however, none of them are identified as members of the stream according to their criteria. We identified 6 of them as members, and 3 more also have a $p_\text{memb}>0.5$ but they are too far from the celestial track.

Since this track does not have previously reported distance information as a function of $\phi_1$, we fitted a second degree polynomial, obtaining the following distance track shown in the top panel of Fig.~\ref{fig:dist_gradients} as a red dashed line:
\begin{equation}
    d \text{[kpc]} = 27.733 + 4.812\times10^{-1}\phi_1-5.368\times10^{-3}\phi_1^2.
\end{equation}
\noindent \citet{Li2022} using the RRL of the $S^5$ survey in the Turranburra region inferred a mean distance of $26.3$~kpc (black dashed line in the top panel of Fig. \ref{fig:dist_gradients}), consistent with our results.

\subsubsection{Elqui} \label{sc:elqui}

We identified 10 RRL in the stream, expecting 0.7 of them to be contaminants. With them, we inferred a new distance gradient by fitting a 1D polynomial, obtaining the following track shown in the middle panel of Fig. \ref{fig:dist_gradients} as a red dashed line:
\begin{equation}
    d\text{[kpc]}=50.524 + 1.071\phi_1
\end{equation}
\noindent \citet{Li2022} using the RRL of the $S^5$ survey in the Elqui region inferred a gradient distance that is in excellent agreement with ours\footnote{\citet{Li2022} adopted a left-handed coordinate system to define their stream frame of reference; therefore, we invert the sign of the angular coordinate $\phi_1$ in their distance modulus gradient to be consistent with the right-handed coordinate system used here.}
. In addition, all members, except for one already reported as a member by \citetalias{Shipp2019}, are new detections.

The photometric metallicity of the identified members is consistent with that reported by \citet{Li2022}. They have a median of [Fe/H]=$-2.06\pm0.09$~dex with a spread of 0.15~dex. However, there is one star (Gaia~DR3 5000591527509120384) that has a photometric metallicity of [Fe/H]=$-1.75 \pm 0.28$~dex, more metal-rich than the rest of the members but still consistent within the errors.

\subsubsection{Indus} \label{sc:indus}

We identified 4 RRL in the stream (expecting 0.7 contaminants); this allows us to assign a more precise value to the distance track of the Indus stream (bottom panel of Fig. \ref{fig:dist_gradients})
. We fitted a first degree polynomial, obtaining the following distance track: 
\begin{equation}
    d \text{[kpc]} = 14.38 - 8.27\times10^{-2}\phi_1.
\end{equation}
\noindent \citet{Li2022}'s distance gradient is consistent with ours\footnote{We noticed that the slope of the reported distance gradient of Indus in \citet{Li2022} is flipped with respect to their reported members. In Fig. \ref{fig:dist_gradients} we show it corrected.}. The subtle difference between both tracks can be explained by the assumptions taken for the metallicities; an error of 0.32~dex (typical metallicity error) implies an error of 0.10 mag in absolute magnitude \citep[from the absolute magnitude-distance relation,][]{Muraveva2018}, which -- at the distance of Indus ($\sim16.6$~kpc) -- implies an error of $\sim0.78$~kpc.

\subsubsection{Phlegethon}
We identified 3 RRL in the tails; these are new detections and have consistent photometric metallicity with the stream. One of them lies on the extended track of the stream, verifying that this extrapolation is correct.

\subsubsection{Remaining streams without an associated progenitor} 

The other streams without an associated globular cluster studied in the present work are: Gaia1, Gaia6, Gaia10, Gaia12, Jet, Leiptr, Ophiuchus, Sylgr, and Ylgr. These streams have between 1 and 2 RRL; therefore, we restrict ourselves to giving an average measurement of the distance (see Table \ref{tb:resumen_results}) since, with so few RRL, it is not possible to perform a statistical analysis of the population.

\section{Discussion} \label{sc:discussion}

\subsection{Segregations and bimodalities}

After analyzing all the 56 stellar streams with proper motion tracks and distance measurements available in \verb|galstreams|, we found a segregation by RRL type exclusively in the Pal~5 stream. Resolving whether this kind of segregation was a common feature between the streams or just a peculiarity of Pal~5 was one of the motivations of this work. While it seems to be one of a kind, we did observe a segregation by Oosterhoff type in three streams: Cetus-Palca, AAU and M92. In the Cetus-Palca stream, while the majority of their RRab members are OoII (39 of 52), we identified a concentration of OoI RRL (9 of 13) at one of the extremes of the reported track; however, this region is close to the LMC in the sky and one of them was identified as an LMC member by \citet{Soszynski2016}; therefore the rest of the OoI RRL of this group might be contaminants that belong to the LMC too. In the AAU stream, we identified 8 RRab, of which 3 are OoI and 5 are OoII; since the spread in metallicity of this stream is small \citep{Li2021}, this segregation might be due to stellar evolution where the OoII are leaving the zero-age horizontal branch, increasing their luminosity and therefore their period. In the M92 stream, there is a clear segregation between the RRL of the cluster and tails, where the cluster ones are OoII and the tails ones are OoI; in this case, this is a consequence of the different metallicity of both groups of RRL \citep[lower metallicities imply higher pulsation periods][]{Catelan2015}, as described below.

Moreover, we observed three cases with a bimodality in metallicity: $\omega$Cen, Jhelum and M92. In $\omega$Cen, we observed a bimodal segregation in the RRL population of the cluster ([Fe/H] = -2.11 and -1.50~dex) consistent with the multiple populations reported using other tracers \citep[{[Fe/H]} = -2.11, -1.85, -1.50 and -1.19~dex,][]{Kuzma2025}; we also identified these two populations in the RRL population of the tidal tails. In the case of Jhelum, we found that its narrow and broad components have different distances and metallicities: while the narrow one is more distant ($d=12.49 \pm 0.36$~kpc) and more metal-poor ([Fe/H]= $-2.23\pm0.13$~dex), the broad one is closer ($d=10.86 \pm 0.32$~kpc) and more metal-rich ([Fe/H]= $-1.74\pm0.14$~dex); this supports the scenario where Jhelum is the result of both a disrupted globular cluster and dwarf galaxy that were accreted by the Milky Way. Finally, in the case of M92, a bimodality in metallicity with a difference of $\sim$0.6~dex is observed between the RRL population of the cluster ([Fe/H]=$-2.2\pm0.1$~dex) and the tidal tails ([Fe/H]=$-1.6\pm0.1$~dex); this, along with other evidences, leads us to hypothesize that M92 is the nuclear star cluster of a disrupted dwarf galaxy, which is the one that forms the stream, similar to M54 and Sagittarius, or LMS-1 and NGC 5053 and NGC 5054.

In summary, the segregation of the RRL in the streams by type of RRL or Oosterhoff is not a standard feature along the streams; but it is rather a peculiarity of a few, or even contamination from other substructures. On the other hand, the presence of segregation and bimodalities in metallicity indicates a more complex origin history, either because the progenitor of the stream may be a dwarf galaxy, or because it has more than one associated component, such as a globular cluster embedded in its host galaxy that may or may not be both disrupted.

\subsection{RRL in the progenitor and its tidal tails} \label{sc:discussion2}
A relevant observation from our analysis of the RRL in the clusters and tails of streams with an associated progenitor is that those whose progenitor has a significant number of RRL, have only a handful in their tails, such as in M3, $\omega$ Cen, or NGC 3201. This can be explained through mass segregation.

Mass segregation is a well-known dynamical process in globular clusters, whereby the more massive members gradually migrate to the cluster center, while the lighter ones tend to move farther from it \citep{Spitzer1987}. To study the impact of this process on stellar escape and tidal tail formation, \citet{Balbinot2018} applied the cluster evolution code \verb|EMACSS| \citep{Gieles2014,Alexander2014}, combined with a semi-analytic model for the evolution of the stellar mass function, and found that the first stars to escape are preferentially of low mass. 
As a result, stellar streams are more easily detected when the progenitor is close to dissolution, such that a higher fraction of massive (and therefore brighter) stars are in the tails; and near apocenter, where the stream is in the most densely packed stage.

The RRL, being evolved stars, are among the most massive stars that are still alive in the population; %
thus, their escape is a sign that the progenitor of the stream is in the final stages of dissolution, as seen in Pal 5. Vice versa, this implies that progenitors with the largest populations of RRL are not necessarily optimal targets for identifying RRL in their tidal tails or to detect the tails at all.

\begin{figure}[h!]
    \centering
        \includegraphics[width=\linewidth]{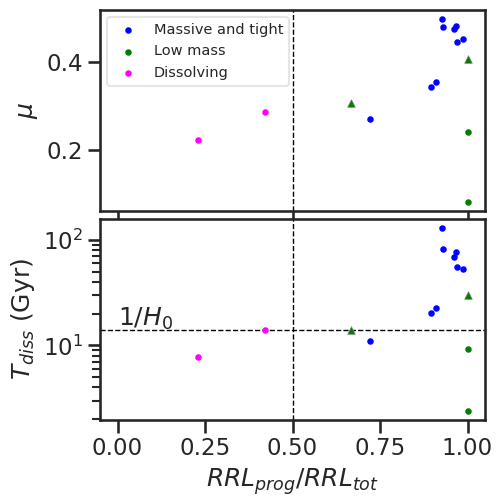}
    \caption{Remaining mass fraction of the cluster ($\mu$, top panel) and lifetime ($T_\text{diss}$, bottom panel) as function of the ratio between RRL on the progenitor and the total. The blue dots correspond to streams with many RRL ($>$10) and the majority of them ($>$50\%) in the cluster (massive and tight), the green symbols correspond to those with few RRL and the majority of them in the cluster (low mass); and the magenta dots to those with the majority of the RRL in the tidal tails (dissolving). In the low mass group, the triangle symbols correspond to NGC 1261 and NGC 2808 which have more RRL than the ones we identify \citep[$>$10,][]{Reyes2024}; therefore, actually belonging to the massive and tight group.}\label{fig:discussion1}
\end{figure}

In Fig. \ref{fig:discussion1}, we can empirically see this, where we show the remaining mass fraction of the cluster ($\mu=M_{\text{prog}}/M_{\text{tot,in}}$, top panel) and lifetime ($T_\text{diss}$, bottom panel) as a function of the ratio between RRL on the progenitor and the total. Both parameters, $\mu$ and $T_\text{diss}$, were estimated by \citet{Baumgardt2019} and are available in the DbGC\footnote{The present-day masses, $M_\text{prog}$, were derived by \citet{Baumgardt&Hilker2018} and \citet{Baumgardt2019} by fitting N-body dynamical models to velocity-dispersion profiles based on Gaia DR2 proper motions and line-of-sight velocities. The initial masses, $M_\text{tot,in}$, were inferred by integrating cluster orbits backwards in time and modeling mass loss due to dynamical evolution, which also provided estimates of the dissolution timescales, $T_\text{diss}$.}. We separate the streams with an associated progenitor of our sample into three groups: 1) Massive and tight: those with many RRL ($>$10) and the majority of them ($>$50\%) in the cluster (M3-Svöl, $\omega$Cen-Fimbulthul, NGC 3201-Gjöll, M5, M68-Fjörm, NGC 5466, M2, NGC 1851, and NGC 6101) shown as blue dots; 2) Low mass: those with few RRL and the majority of them in the cluster (NGC 2298, NGC 1261, NGC 288, NGC 2808, and NGC 6397)\footnote{NGC 6397 has zero RRL at all and, therefore, is not represented in Fig. \ref{fig:discussion1}.} shown as green symbols; and 3) Dissolving: those with the majority of the RRL in the tidal tails (Pal 5 and M92 \footnote{In the case of M92, according to our hypothesis (see Sec. \ref{sc:M92}), the RRL of the cluster and tails are independent of each other; thus, this stream would not formally belong to the third group, but we keep it in it.}) shown as magenta dots. These numbers indicate that streams with the majority of the RRL in their tails are not common, suggesting that this might be a short stage in the process of producing a stellar stream.

This classification is consistent with the lifetimes of their progenitors \citep{Baumgardt2019,Baumgardt2023} and with the expected outcome from mass segregation. Clusters in the first group have long lifetimes, exceeding a Hubble time in almost all cases, whereas those in the second and third groups with low-mass systems and most of their RRL in their tidal tails, respectively, have short lifetimes. In particular, in the second group, NGC 2298 has a lifetime of only $T_{\text{diss}}=2.4$ Gyr and a remaining mass cluster fraction $\mu=0.08$, but this is a result of its unusually high inferred initial mass \citep{Baumgardt2019,Baumgardt2023}. On the other hand, NGC 1261 and NGC 2808, shown as green triangles in Fig.~\ref{fig:discussion1}, have more than 10 RRL in their progenitors according to \citet{Reyes2024}; therefore, they actually belong to the first group. The relation between the fraction of RRL and mass retained in the progenitor further supports this picture. Clusters with a higher remaining mass cluster fraction tend to retain more of their RRL. 
This is consistent with the expected production rate of these stars and indicates that the RRL escape the progenitor along with an amount of mass proportional to their number. 

We also observed that streams where most RRL are found in the tails tend to lie near apocenter, where tidal debris is most spatially concentrated and easier to detect, although this does not necessarily reflect a direct causal connection with RRL escape, rather an observational bias. Other orbital parameters, including tidal radius, eccentricity, pericenter, and apocenter distance, were explored but showed no clear correlation with the RRL ratio (not shown).

We found that there is a higher fraction of streams with associated RRL in those with progenitors (13/17, $\sim$75\%) than in those without (19/34, $\sim$45\%); this does not contradict what we stated in this section. Even though we expect to find more RRL in the tidal tails of streams whose progenitor is close to dissolution, the absence of RRL in many progenitor-less streams may instead indicate that their progenitors were relatively low-mass systems. Since the number of RRL roughly correlates with the mass of the progenitor, a small number (or absence) of RRL is consistent with low-mass progenitors. Such systems are more susceptible to tidal disruption, making it unsurprising that their progenitors are no longer detectable.

Overall, these results show that the presence of RRL in tidal tails is linked to the late stages of cluster dissolution. Streams with an associated progenitor whose RRL are mostly in the tails are rare and likely represent a brief evolutionary stage, while the majority of systems retain nearly all of their RRL in the progenitor, consistent with the predictions of mass segregation.

\subsection{Intrinsic dispersions}

In this work, we have consistently obtained intrinsic proper motion dispersions, as described in Sec. \ref{sc:result}, for 70 of 75 tracks with proper motion information in the \verb|galstreams| library.

The dispersions that we found might appear large if we compare them with \citet{Li2022}, who, after analyzing a dozen stellar streams with $S^5$ data, concluded that globular cluster streams have velocity dispersions lower than $\sim5$ km/s. However, there are several independent examples of robust cases with high velocity dispersions that are consistent with our results, such as Balasubramaniam et al. (in prep., private communication), who found that the Turranburra and Elqui streams have large dispersions (>20 km/s)\footnote{Both are thought to have a dwarf galaxy progenitor \citep{Shipp2018,Li2022}, which is consistent with these velocity dispersions.}. In addition, theoretical studies, like \citet{Errani2015} and \citet{Penarubia2025}, argue that dwarf galaxies embedded in a dark matter halo can be affected by the dark matter profile or by interactions with dark matter subhalos, heating up the system and inflating their velocity dispersions. On the other hand, it is important to note that the intrinsic dispersions we infer depend on the assumption that the observational uncertainties are correctly estimated. In particular, an underestimation of proper motion and distance errors could artificially inflate the inferred intrinsic dispersion, although this is unlikely given the precision of the Gaia data and the robustness of the distance estimates based on RRL.

\begin{figure}
    \centering
    \includegraphics[width=\linewidth]{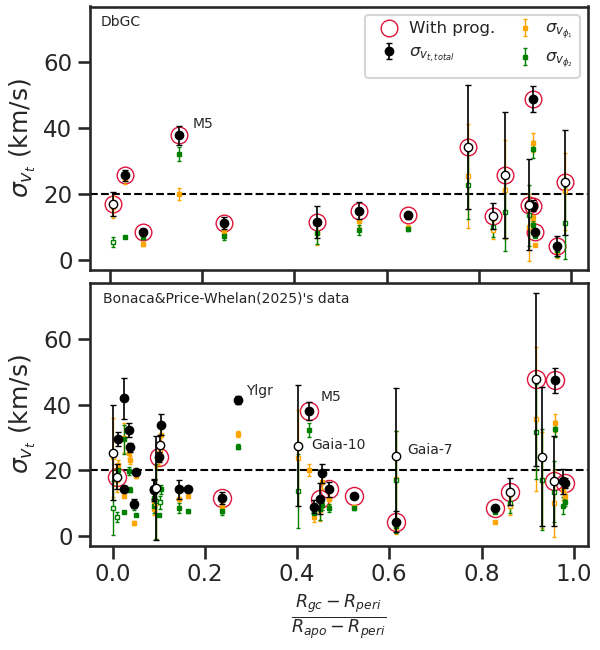}
    \caption{Intrinsic tangential velocity dispersion of the streams as function of their orbital phase. \textbf{Top:} The 17 streams studied with an associated progenitor which orbital phase information is obtained from the DbGC. \textbf{Bottom:} The 39 streams studied with physically consist orbital phase information reported in \citet{Bonaca2025}. The orange and green squares are the velocity dispersions in the components along and perpendicular to the stream, respectively; and the black circles are total velocity dispersions. Fill and empty symbols are the streams with and without members already reported in the literature, respectively; and therefore, with and without prior information of their intrinsic tangential velocity dispersion. The red empty circles indicates the streams associated with a progenitor.}\label{fig:sig_v}
\end{figure}

Moreover, \citet{Helmi1999} showed through numerical simulations that the velocity dispersion of stellar streams can increase significantly at orbital turning points (apocentres or pericentres)\footnote{Liouville's theorem.}. In Fig.~\ref{fig:sig_v}, we present the intrinsic tangential velocity dispersion of the streams inferred in this work, $\sigma_{v_t}$, as a function of their orbital phase. The top panel shows the 17 streams with associated progenitors, where the orbital phase information was taken from the DbGC (see Sec.~\ref{sc:data}), and the tangential velocity dispersions were computed using the median stream distances reported in \verb|galstreams|. The bottom panel displays 39 of the 57 analyzed streams with physically consistent orbital phase information ($0 < \frac{R_{\text{gc}} - R_{\text{peri}}}{R_{\text{apo}} - R_{\text{peri}}} < 1$) reported by \citet{Bonaca2025}; the tangential velocity dispersions were calculated using the mean stream distances also reported by \citet{Bonaca2025}.

Among the streams shown in the top panel, 6 out of 17 present velocity dispersions higher than 20~km/s, all located near orbital turning points. Similarly, in the bottom panel, 16 out of 39 streams show high dispersions, 75\% of which are also found near turning points. The correlation between high dispersions and orbital turning points is consistent with the results of \citet{Helmi1999} and indicates that these high inferred dispersions might be an actual dynamical feature. Additionally, in the streams that have members already reported in the literature (filled symbols in Fig. \ref{fig:sig_v}), we used them to obtain a reference value of the intrinsic dispersions, which then served as the prior for our RRL-based inference (see Sec. \ref{sc:priors}). The dispersions inferred from these reported members were already large, and our results are consistent with them. 

The main difference between the two panels of Fig. \ref{fig:sig_v} is the orbital phase of the streams, which strongly depends on the assumed Galactic potential and the accuracy of the orbital information of the object. Since the streams with progenitors have more robust orbital information, DbGC's orbital phases are more trustworthy; however, \citet{Bonaca2025} do not limit themselves to streams with a surviving progenitor. In particular, we highlight the case of M5, where using the orbital phase of 0.43 reported by \citet{Bonaca2025} would result in an anomalously high dispersion. In contrast, adopting the -- more reliable -- orbital phase of 0.15 reported in the DbGC yields a dispersion consistent with \citet{Helmi1999} results. A similar situation might also be occurring in the other three anomalous cases: Gaia–7, Gaia–10, and Ylgr.

In summary, while our intrinsic dispersions may appear large, there are several studies, both theoretical and observational, that are consistent with our results. The ones that appear unusually large ($>$20 km/s) correspond to streams that are located at the turning points of their orbits, consistent with \citet{Helmi1999} predictions.

\section{Conclusions} \label{sc:conclu}

In this work, we have presented the first systematic census of RRL stars kinematically associated with known stellar streams, motivated by the need for a homogeneous and distance-precise tracer that enables a full characterization of the streams. We constructed a probabilistic membership model based on proper motion and distance information, which identifies the RRL associated with  known streams with proper motion reported in \verb|galstreams|, and used these RRL to derive distance gradients (or anchor mean distances) as a function of an angular coordinate along the stream, to establish their intrinsic dispersions, and to study and compare the RRL population of the streams and their progenitors.

For our search we used a compilation of the largest Milky Way RRL catalogs (Gaia SOS, PS1, and ASAS-SN-II), with $\sim2.8\times10^5$ RRL after quality cuts. Among the 56 streams (75 tracks) that we studied, 17 have an associated progenitor and 39 do not; there were 12 streams with more than one track reported in the literature. Of those with a surviving progenitor, we identified RRL in the tails of 13 stellar streams, 5 of them with more than 3 RRL. Of those with no progenitor, we identified RRL in 20 stellar streams, 9 of them with more than 3 RRL. We recover and expand previously known RRL associations, identifying new candidate members; except in the case of the GD-1 stream, in all the other 32 streams where we identify RRL, we detect at least one new member. We summarize the properties of the tracks in Tables~\ref{tb:resumen_results} and give a list of the candidate members for each track in Table~\ref{tb:membs}.

Another important result of the census is that in all the streams with identified RRL we were able to anchor more precise distances to streams with only a mean value, confirm already reported distance gradients, and, in particular, infer new distance gradients to 5 streams: both components of Jhelum, LMS-1, Turranburra, Indus, and Elqui. This improved distance information is crucial to make full phase-space studies of the stellar streams themselves and of the Galaxy, such as inferring the Galactic potential from their orbits, studying the nature of the dark matter, or rebuilding the assembly history of the Milky Way.

From a global perspective, the census discards some particular patterns and reveals others.
\begin{enumerate}
    \item The segregation by RRL type reported in the Pal~5 stream by \citet{Price-Whelan2019} was recovered. A similar effect is not observed in any of the other streams; thus, it is safe to conclude that this is a peculiarity of this particular stream. While this could be attributed to low-number statistics, its true origin remains elusive.  
    \item Particular segregations by Oosterhoff were observed in three streams; in one case, it is likely contamination from the LMC (Cetus-Palca), in another it might be due to the RRL leaving the zero-age horizontal branch (AAU), and in the other it is due to a bimodality in the metallicity of the RRL population (M92). 
    \item The bimodality in metallicity observed in three streams ($\omega$Cen, Jhelum, and M92) reflects a complex origin story; in particular, for the case of Jhelum, this supports the scenario where the stream is the result of a globular cluster (narrow component) and a dwarf galaxy (broad component) accreted and disrupted by the Galaxy. For the case of M92, we propose the scenario in which M92 is the nuclear star cluster of a disrupted host dwarf galaxy responsible for the stream observed by \citetalias{Ibata2021}; meanwhile, the \citet{Sollima2020}'s and \citetalias{Thomas2020}'s tracks might trace the tidal tails of the cluster; to confirm these hypotheses, more information and analysis are needed.
    \item Streams with an associated progenitor that have a large fraction of their RRL in the tidal tails are uncommon and appear to trace progenitors in advanced stages of dissolution (e.g., Pal~5), consistent with expectations from mass segregation and cluster evolution \citep{Balbinot2018}. 
    \item We have consistently obtained, for the first time, intrinsic velocity dispersions for 70 of the 75 tracks with proper motion information in the \verb|galstreams| library. This is a relevant contribution of this work, as it is required by automated searches, such as the census conducted here, and it should facilitate similar future studies not limited to RRL stars. This information will be added to \verb|galstreams| in its next version. 
    \item The intrinsic velocity dispersions inferred for several streams are large \citep{Li2022}; these large dispersions correlate in many cases with orbital turning points \citep{Helmi1999} and with previously reported high-dispersion samples, suggesting that some measured dispersions might be real dynamical features; however, reliable inference requires larger tracer samples than the small-number statistics of the RRL. 
\end{enumerate}

In summary, this first systematic RRL census across all known streams with reported distances and proper motions provides a step forward in a homogeneous characterization of the Milky Way tidal debris. These results expand the knowledge of the known (RRL) members in many streams, anchor precise distances, and reveal patterns in RRL populations that shed light on the progenitor lifetimes and the stream formation scenarios. The method presented here can be applied to other types of standard candles, and the catalog of high-probability members can be combined with spectroscopic follow-up to obtain full phase-space information. Such combined analyzes will allow for more reliable inference of stream properties and improved constraints on the Galactic potential.

The experience and results of the census also leave us with lessons on aspects that can be improved. We highlight the sensitivity of the method to the choice of tracks. In more than one case, an identified RRL was classified as a member according to one track but not by others. 
In addition, some streams associated with a progenitor have reported tracks in the literature that show mismatches in sky position (e.g., $\omega$~Cen), distance (e.g., M68–Fjörm), and/or proper motion (e.g., NGC~1851) relative to their progenitor. This census represents a first approach to find out which of the studied streams contain a significant number of RRL based on the currently reported tracks. In future work, these issues could be addressed by running the inference blindly where possible and, in addition, inferring new RRL–based tracks for streams with sufficient members, as well as for those lacking any prior information.

\begin{acknowledgements}
The authors would like to acknowledge Alex Riley, Nora Shipp, Ting Li, Julio Carballo-Bello and Jorge Peñarrubia for useful comments and relevant insights for the discussion; as well as Zhen Yuan and Akshara Viswanathan for kindly sharing supplementary data that allowed further comparison of our results. BD and CM thank Adrian Price-Whelan for his invitation to the Flatiron Institute, where useful discussions took place. 

The author BD gratefully acknowledges the scholarship Becas de apoyo a docentes para estudios de posgrado en la UdelaR, Maestría, 2023,
from the Comisión Académica de Posgrado (CAP) for financially supporting
and making this work possible.

This research was partly supported by funding from the Programa de Desarrollo de Ciencias Básicas (PEDECIBA), the MIA program and project grant C120-347 of the Comisión Sectorial de Investigación Científica (CSIC) at Universidad de la República, Uruguay and project grant FCE\_1\_2021\_1\_167524 from the Fondo Clemente Estable of the Agencia Nacional de Innovación e Investigación (ANII).

This research made use of the high-performance computing facilities
of ClusterUY (Nesmachnow \& Iturriaga 2019), for which the authors are thankful. 

GFT acknowledges support from the Agencia Estatal de Investigación del Ministerio de Ciencia en Innovación (AEI-MCIN) under grant number PID2023-150319NB-C21 and the grant RYC2024-051016-I funded by MCIN/AEI/10.13039/501100011033 and by the European Social Fund Plus. 

Finally, BD warmly thanks Mauro Cabrera, Rafael Bertolotto, Rodrigo Cabral, Juan José Downes, and Selena Seidel for closely following this work from its early stages and for sharing their wisdom and insights, some of which were key to the development of this research. \\

\\
{\it Software:} 
    \textsc{emcee} \citep{Foreman-Mackey2013}, 
    \textsc{astropy} \citep{astropy2018},
    \textsc{gala} \citep{gala},
    \textsc{matplotlib} \citep{mpl},
    \textsc{numpy} \citep{numpy},
    \textsc{scipy} \citep{scipy2001},
    \textsc{jupyter} \citep{jupyter2016}, and 
    \textsc{topcat} \citep{Topcat2005,Stilts2006}

\end{acknowledgements}

\bibliographystyle{aa} 
\bibliography{refs,softrefs}

\begin{appendix}

\section{Details of the methodology} \label{ap:method}

\subsection{Tracks with particular sample selection}

In some tracks, we took more extended regions on and off-stream in $\phi_2$ in order to have enough stars for a robust representation of the background. In the case of TucanaIII-S19 \citepalias{Shipp2019}, we took a width of $9\sigma_{\phi_2}$ for each region (and therefore add/subtract $15.5\sigma_{\phi_2}$ to define the window's width). For M2-I21 and Kshir-I21 \citepalias{Ibata2021}, we took $8\sigma_{\phi_2}$for each region ($14\sigma_{\phi_2}$ to define the window's width). For Elqui-S19 \citepalias{Shipp2019} and Spectre-C22 \citep{Chandra2022}, we took $10\sigma_{\phi_2}$ and $8\sigma_{\phi_2}$, respectively, for the off-stream region (therefore, $15\sigma_{\phi_2}$ and $13\sigma_{\phi_2}$ to define the window's width, respectively).

\subsection{Bayesian model details} \label{ap:bayes}

We constructed the Bayesian probabilistic mixture model following \citet{Price-Whelan2019}, which takes the form:
\begin{equation} \label{eq:posterior}
    p\left(\mathbb{C}_{int}, f\ |\ \vec{y_n}\right) \propto p(\mathbb{C}_{int}, f)\prod_n^N  \mathcal{L}(\vec{y_n}\ |\ \mathbb{C}_{int},f),
\end{equation}

\noindent where $p\left(\mathbb{C}_{int}, f\ |\ \vec{y_n}\right)$ is the posterior probability of the model parameters $\mathbb{C}_{int}$, the intrinsic covariance matrix, and $f$, the stream's RRL fraction; $p(\mathbb{C}_{int}, f)$ is the prior probability distribution over the parameters, and  $\mathcal{L}(\vec{y_n}\ |\ \mathbb{C}_{int},f)$ is the likelihood probability for the $n$-th star, given the model parameters. 

\subsubsection{The likelihood}

The likelihood is modeled as the weighted sum of the probability that the RRL belongs to the stream, $p_{\text{st}}$, plus the probability of belonging to the background, $p_{\text{bkg}}$: 
\begin{equation} \label{eq:likelihood}
    \mathcal{L}(\vec{y_n}\ |\ \mathbb{C}_{int}, f) = f p_{\text{st},n} + (1-f)p_{\text{bkg},n} 
\end{equation}
\noindent The weight $f$ is the stream versus background RRL ratio in the on-stream region (i.e., where the inference is made) and is a free parameter of the model.

\paragraph{Stream probability}
\ \\

The probability that a star belongs to the stream $p_{\text{st},n}=\mathcal{N}(\vec{y_n}\ |\ \vec{y}(\phi_1),\mathbb{C})$ is modeled as a multivariate normal distribution, whose center follows the stream track\footnote{Internally, the \texttt{galstreams} track is fitted in $(\phi_1,\phi_2)$ with a third degree polynomial to provide the extrapolation outside the reported track and for the likelihood to be evaluated at any arbitrary $\phi_1$.}
$\vec{y}(\phi_1)$. The Gaussian noise, $\mathbb{C}$, is taken as the quadratic sum of the intrinsic dispersion $(\mathbb{C}_{int})$ and observational noise $(\mathbb{C}_{obs})$ in proper motion and distance, where the intrinsic dispersion is a diagonal matrix consisting of $\left(\sigma^{int}_{\mu_{\phi_1}}, \sigma^{int}_{\mu_{\phi_2}}, \sigma^{int}_d\right)$ squared. These are, in addition to $f$, the remaining free model parameters.

\paragraph{Background probability}
\ \\

We calculated the probability that a star belongs to the background, $p_{\text{bkg}}$, from a fixed empirical Gaussian mixture model of the proper motions and distance inferred from the RRL in the off-stream region. %
We assumed here that the proper motion and distance are locally independent of their position in the sky, which allows us to make cuts in the sky without biasing the sample. Under this assumption, the distribution of proper motions and distances that we obtain in the off-stream region can be considered representative of the on-stream region.

The number of Gaussians used in the mixture model is determined by choosing the number of components that minimizes the Bayesian Information Criterion (BIC). Additionally, after some trial and error, we determined that, on average, there should be more than 15 RRL per Gaussian so that the background is well modeled and the results can be trusted. The streams whose backgrounds do not achieve this condition cannot be reliably studied using this method (see Sec. \ref{sc:result_summary}).

\subsubsection{The priors} \label{sc:priors}
We assume that the prior is separable in all parameters. For the weight $f$, we adopt a uniform distribution between $0$ and~$1$. 
For the prior on the intrinsic dispersions (in proper motions and distance), we use a hierarchical inference model. For the proper motion dispersions, we first infer the individual intrinsic tangential velocity dispersion for each stream with published members (generic stars, non-RRL) by de-convolving the observational errors. The resulting distribution of inferred intrinsic dispersions is used as a hyper-prior probability distribution for the \emph{population} of streams, which is then used as a prior for the inference of the intrinsic proper motion dispersions for each individual stream based on the RRL. For the distance dispersions, we used a uniform prior between 0.001~kpc and 5~kpc as a hyper-prior.

\paragraph{Intrinsic proper motion dispersion hyper-prior}\ \\

We first created a deconvolved distribution of the tangential velocity dispersion. To do this, we need proper motions, distances, and their uncertainties. Unfortunately, of the $\sim$150 streams, only 4 have all of this data for RRLs\footnote{NGC5466-J21 \citep{Jensen2021}, Pal5-PW19 \citep{Price-Whelan2019}, Cetus-Palca-T21 \citep{Thomas2022}, and LMS1-M21 \citep{Malhan2021}}; therefore, we looked for additional sources of information. We used the streams from \citetalias{Ibata2021} that have reported members (in general, not restricted to RRL) with both proper motion and distance. Proper motion uncertainties were obtained from Gaia~DR3. For most reported members, distance uncertainties are not provided. To estimate them, we selected the \citetalias{Ibata2021}'s streams with a living progenitor and reported members within the progenitor's tidal radius and assumed that the true size of the progenitor is negligible compared to its distance and, therefore, that the distance dispersion is solely due to uncertainties. Based on these estimates, the distance uncertainties were found to be $\sim10$\%.

Applying this distance error to the other streams with members and distance data from \citetalias{Ibata2021}, we calculated their tangential velocities along with their respective uncertainties. Using an extreme deconvolution model \citep{Holoien2017} with a single Gaussian, a tolerance of $1\times10^{-8}$ and a maximum of iteration of 2048, we obtained a deconvolved dispersion of tangential velocities for each stream. 
We then fitted a gamma distribution to the resulting distribution of tangential velocity dispersions of the \emph{population} of streams, for which we obtained $\alpha = 1.97507$ and $\beta = 0.06927$. This result was obtained using only streams with reported members from \citetalias{Ibata2021}; nevertheless, similar values are found when the distance error calculation is extended to include all streams with reported members.

This distribution is then transformed into proper motion space using a median distance for each stream ($d_{st}$) according to:
\begin{equation}\label{eq:priorpm}
    p\left(\sigma^{int}_{\mu_{\phi_i}}\right) = f_{\Gamma}\left(\kappa\sqrt{2}d_{st}\sigma_{\mu_{\phi_i}}^ {int}\right)\left|\kappa\sqrt{2}d_{st}\right|
\end{equation}

\noindent where $\kappa = 4.74$ km~yr/s and with
\begin{equation}
    f_\Gamma (x) =\frac{\beta^\alpha}{\Gamma(\alpha)}x^{\alpha-1}\exp{(-\beta x)},
\end{equation}

\noindent so we can use it as a prior for the intrinsic proper motion dispersions for each stream, $\sigma_{\mu_{\phi_1}}^ {int}$ and $\sigma_{\mu_{\phi_2}}^ {int}$.\footnote{$d_{st}$ in~kpc and $\sigma_{\mu_{\phi_i}}^ {int}$ in mas/yr.}

\paragraph{Streams without reported members}
\ \\

For the case without reported members, the hyper-prior in Eq. \ref{eq:priorpm} inferred for the population of streams is used as the prior in the inference of the intrinsic proper motion dispersions. For the intrinsic distance dispersion, the flat hyper-prior mentioned above is used.

\paragraph{Streams with reported members}
\ \\

For the case with reported members, we used the hyper-priors and the published stream members to obtain a posterior probability, which is then used as the (more informative) prior in the inference with the RRL.

To obtain the posterior probability, we used a simple Bayesian model where the likelihood is a Gaussian distribution, $\mathcal{N}(y_{i,n}~|~y_i(\phi_1), \mathbb{C}_i)$, centered on the track $y_i(\phi_1)$, with the covariance matrix being the sum of the observational and intrinsic noise, $\mathbb{C}_i = {\sigma^{obs}_{i,n}}^2 + {\sigma^{int}_{i}}^2$, the latter being the free parameter we want to infer. To sample the posterior probability, we used the same Markov Chain Monte Carlo (MCMC) sampler described in Sec. \ref{sc:MCMMC}. However, since this inference involves only one free parameter, we ran the sampler with 8 walkers. This posterior has a mode and a variance, which are used to calculate the parameters $\alpha$ and $\beta$ of a gamma distribution\footnote{A gamma distribution satisfies: mode = $(\alpha-1)/\beta$ and $\sigma^2=\alpha/\beta^2$} that serves as a bespoke prior for each stream with reported members.

For the posterior probability of the intrinsic distance dispersion, we have three cases of reported members: \textit{i)} members with distances and their uncertainties, \textit{ii)} members with distances but without their uncertainties, and \textit{iii)} members without distances nor their uncertainties. In the first case, we use the data from the members. In the second case, as we do not have the uncertainties available, we used the photogeometric distance uncertainties from Gaia~DR3. For the last case, as we do not have any distance information, the flat hyper-prior was used. In addition, there were some cases in which the inference based on the reported members did not converge; in this case the flat hyper-prior was used instead.

\subsection{Markov Chain Monte Carlo sampler} \label{sc:MCMMC}

We used the MCMC affine-invariant ensemble sampler \verb|emcee| \citep{Foreman-Mackey2013} to generate samples of the posterior probability given by Eq. \ref{eq:posterior} over the parameters $(\mathbb{C}_{int}{\scriptstyle(\sigma^{int}_{\mu_{\phi_1}}, \sigma^{int}_{\mu_{\phi_2}}, \sigma^{int}_d)}, f)$. The inference is made using only the RRL within the stream region. As we have 4 free parameters, we used 40 
walkers to run \texttt{emcee} for an initial $2^9$ steps to `burn-in' the sample, then reset and restart the sampler for another $2^{14}$ steps. We thinned the chains by keeping only every 225$^\text{th}$ steps, leaving us with $M=2880$ posterior samples. The thinning value is chosen so that it is greater than the autocorrelation time (estimated with \verb|emcee|) for all streams.

\section{Streams with associated progenitor and 1-2 RRL in tails}

\subsection{M68-Fjörm} 

There are three 5D tracks available on \verb|galstreams| for this stream: M68-P19 \citep[default,][]{Palau2019}, Fjorm-I21 and M68-I21  \citepalias{Ibata2021}. However, the latter track has several discrepancies with the M68 cluster and the other two tracks; therefore, it is not taken as a track for this stream \citep[as suggested by][]{Mateu2023}, but for a completely different stream without a progenitor associated. The other two tracks are in disagreement with the cluster's distance and between them at $\phi_1\lesssim30^\circ$, with \citetalias{Ibata2021} predicting a shorter distance than \citet{Palau2019}, as shown in the fifth row of Fig. 11 from \citet{Mateu2023}.  

Using the default track \citep{Palau2019}, with or without applying a distance correction to align it with the cluster, we identified all 33 RRL within the tidal radius that have RUWE$<$1.4 \citep{Reyes2024} and only a single RRL in the tails. This is another case, like with NGC 3201 and $\omega$Cen, where we observe a significant amount of RRL in the progenitor and only a handful in its tidal tails; given the mass segregation that occur in globular clusters this result is not unexpected (see discussion in Sec. \ref{sc:discussion2}).

\subsection{M2 (NGC 7089)}

We detected all the 17 RRL with RUWE$<$1.4 within the cluster in our sample \citep{Reyes2024} and 2 RRL in the tails that are new detections. The default track \citep{Grillmair2022} does not have distance information; therefore, to apply the method we took the shorter distance track from \citetalias{Ibata2021} and extrapolated it along $\phi_1$. The two RRL identified in the tails are in the segment of the stream detected by \citetalias{Ibata2021}, therefore, we cannot confirm whether the extrapolation made of the distance track was correct.

\subsection{M5 (NGC 5904)}
Using the default track in \verb|galstreams|, the one by \citet{Grillmair2019}, we detected 71 RRL within the cluster (6 more than the ones identified by \citet{Reyes2024}) and 1 in the tails, which is a new detection. \citet{Abbas2021} associated 6 RRL to the stream, 3 with high confidence, 1 with intermediate, and 2 with low. We detected one of the high-confidence ones within the tidal radius of the cluster, the other 2 have $p_\text{memb}>0.5$ but they are too far from the celestial track ($6.09\sigma_{\phi_2}$ and $10.17\sigma_{\phi_2}$). Using, instead, the track from \citetalias{Ibata2021} we identified 2 more RRL in the tails, one is the intermediate-confidence RRL from \citet{Abbas2021} and the other is already detected with the default track, but again too far from the sky track ($2.94\sigma_{\phi_2}$). This once again highlights the sensitivity of the method to the tracks used. The other two low-confidence RRL from \citet{Abbas2021} are rejected by our model by us because of the large differences in proper motion with respect to the track.

\subsection{NGC 1261}
We identified 2 of the 10 RRL with RUWE$<$1.4 within the cluster in our sample \citep{Reyes2024} and 1 RRL in the tails; however, because of its low metallicity \citep[{[Fe/H] = $–2.68 \pm 0.33$~dex, }][]{Li2023} it is likely a contaminant. \citet{Abbas2021} associated 2 RRL with the NGC 1261 stream, one with intermediate confidence and one with low confidence, we rejected all of them as stream members.

\subsection{NGC 5466}
We identified 20 of the 22 RRL with RUWE$<$1.4 within the cluster in our sample \citep{Reyes2024}; however, we identified 1 RRL as a cluster member that they did not, and 2 of the 3 stars we rejected have proper motions that differ significantly from that of the cluster. We identified 2 RRL in the tidal tails (1 new detection), one of which was already associated with the stream by \citet{Abbas2021} as a high-confidence candidate; the other high-confidence candidate that they associate is a cluster member under our criteria. They associated 2 more RRL with intermediate and low confidence, but these are not detected by us due to large differences in proper motion with respect to the track.

\subsection{NGC 6101}
We identified the 13 RRL with RUWE$<$1.4 within the cluster in our sample \citep{Reyes2024} and 1 RRL in the tails, which is a new detection.

\subsection{NGC 2808}
\citet{Carballo-Bello2018} and \citet{Kundu2021} reported the presence of extra-tidal stars around NGC 2808, while no tidal tail was found by \citet{Sollima2020}. More recently, \citetalias{Ibata2021} detected a $\sim20^\circ$ long tidal tail nearly parallel to the Galactic disk. These tracks present a similar disagreement as the NGC 1851 stream (see Sec. \ref{sc:NGC1851}) with offsets of $\Delta d =~1.25$~kpc, $\Delta\mu_{\phi_1} =~-2.75$ mas/yr, and $\Delta\mu_{\phi_2} =-0.96$~mas/yr. Like in the case of NGC 1851, we applied a manual correction to make the tracks match the cluster. However, in this case, we did not identify any RRL in the tails, with or without correcting the tracks.

\section{NGC 1851 offset tracks} \label{a:offset_track}

\begin{figure}[h!]
    \centering
    \includegraphics[width=0.719\linewidth]{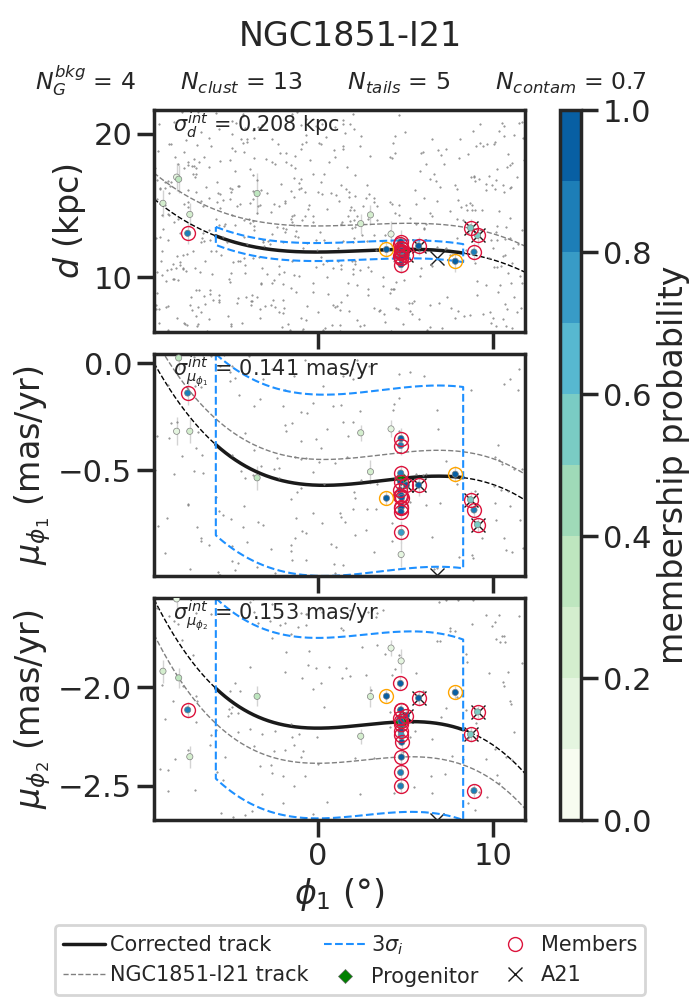}
    \caption{Similar to panels upper left, bottom right and bottom left of Fig. \ref{fig:TucIII_results} but for the NGC 1851 stream. The black crosses are the RRL candidates identified by \citet{Abbas2021}. The gray dashed lines are the original tracks (without correction) from \citetalias{Ibata2021}.
    }\label{fig:NGC1851_results}
\end{figure}

The tracks identified by \citetalias{Ibata2021} for the NGC 1851 stream show an offset with respect to its progenitor in distance and proper motion ($\Delta d=-1.8$~kpc, $\Delta \mu_{\phi_1}=-0.12$ mas/yr, and $\Delta \mu_{\phi_2}=0.18$ mas/yr). We therefore
apply a manual correction to make them match with the cluster before applying our method. The offset is shown in Fig. \ref{fig:NGC1851_results}.  

\section{Turranburra, Elqui and Indus's new distance gradients}\label{a:new_gradients}

\begin{figure}[h!]
    \centering
    \includegraphics[width=0.71\linewidth]{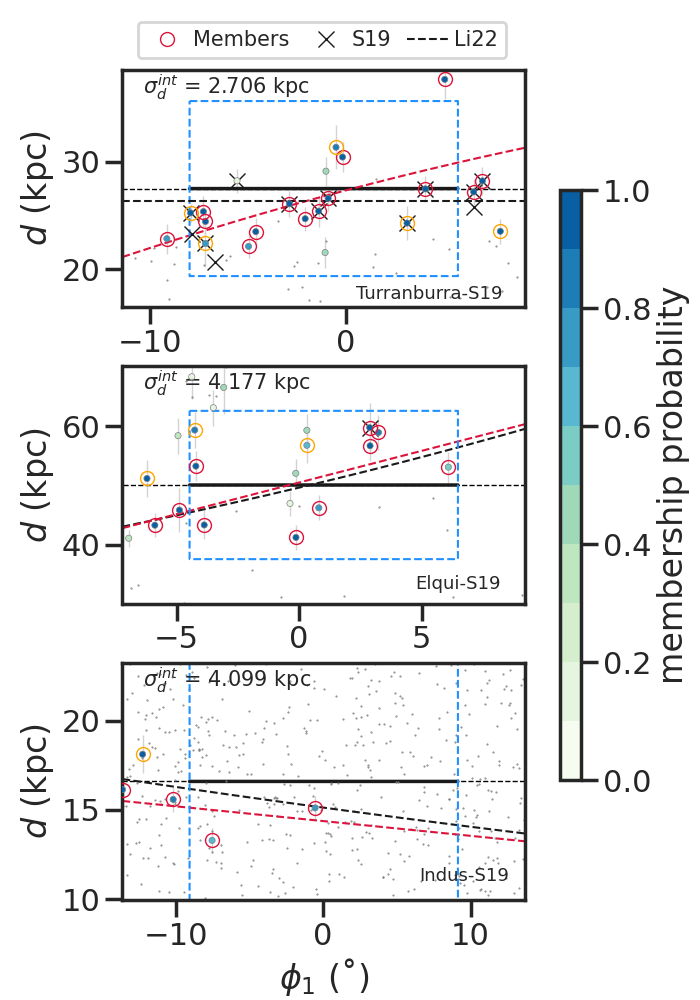}
    \caption{Distance as a function of $\phi_1$ for the Turranburra (top panel), Elqui (middle panel) and Indus (bottom panel) streams. 
    The solid black lines are the mean distances reported by \citetalias{Shipp2019} and the black crosses are the RRL selected by them. The red dashed lines are the distance tracks inferred from the RRL we identified (red circles), and the black dashed lines are the ones proposed by \citet{Li2022}.} \label{fig:dist_gradients}
\end{figure}

The distance gradients of Turranburra, Elqui and Indus inferred in this work, which analytic expressions are in Sec. \ref{sc:turranburra}, \ref{sc:elqui} and \ref{sc:indus} respectively, are shown in Fig. \ref{fig:dist_gradients}.

\section{Properties used for each track}\label{a:track_props}

In Table \ref{tb:resumen_props} we show all the technical properties needed to reproduce our work: the ranges in $\phi_1$, $\phi_2$ and distance of the sample box for each track studied with its reference, the observational width of the stream, the number of RRL in the whole window and in the on and off stream regions, the number of Gaussians used for the background model, the ratio between the extended on-stream region and its complementary area in the sky window, the tidal radius and heliocentric distance of the progenitor when it is still alive, and whether there are previous reported members with proper motion and distance information. The full table can be found in the online version.

\begin{table*}[h!]
    \caption{First 5 rows of a summary of properties used for each track. (1) Name of the track on \texttt{galstreams} (2) Name of the corresponding stream (3) $\phi_1$ range of the box sample (4) $\phi_2$ range of the sample box (5) Distance range of the box sample (6) Sky width from \texttt{galstreams} (7) Whether are members with proper motion information available reported in the literature (8) Whether are members with distance information available reported in the literature (9) Whether the stream has a progenitor associated with (10) Tidal radius of the progenitor (11) Heliocentric distance of the progenitor (12) Total number of RRL in the sample box (\textit{tot}), the on-stream region (\textit{st}), and the off-stream region (\textit{bkg}) (13) Number of Gaussians used for the background model (14) Reference of the track.} \label{tb:resumen_props}
    
    \centerline{
    \resizebox{1.05\textwidth}{!}{%
    \begin{tabular}{cccccccccccccccc}
        \hline\hline
         TrackName & StreamName & $\phi_1$ range & $\phi_2$ range & $d$ range & $\sigma_{\phi_2}$ & pm & $d$ & Prog. & $r_t$ & $d_{\text{prog}}$ & $N$ & $N$ & Ref. \\
         &&$(^\circ)$&$(^\circ)$&(kpc)&$(^\circ)$&avai.&avai.&&(pc)&(kpc)&\textit{\footnotesize{(tot, st, bkg)}}&\textit{\footnotesize{BIC}}&(*) \\
         \hline\\
        Orphan-K19 & Orphan-Chenab & (-156.20, 179.90) & (-10.76, 12.26) & (9.59, 100.00) & 0.8 & T & T & F & - & - & 10263, 1825, 4240 & 9 & (8) \\ 
        Orphan-K23 & Orphan-Chenab & (-104.79, 166.91) & (-10.36, 12.17) & (9.67, 70.00) & 0.8 & T & T & F & - & - & 7883, 1375, 3211 & 8 & (9) \\ 
        Orphan-I21 & Orphan-Chenab & (-64.91, 64.90) & (-8.47, 7.87) & (8.35, 51.29) & 0.56 & T & T & F & - & - & 2563, 442, 1014 & 7 & (7) \\ 
        Chenab-S19 & Orphan-Chenab & (-10.61, 4.46) & (-2.93, 1.98) & (23.88, 55.72) & 0.19 & T & F & F & - & - & 65, 12, 32 & 1 & (16) \\ 
        Cetus-Palca-T21 & Cetus-Palca & (-21.20, 133.99) & (-62.72, 38.05) & (9.83, 50.90) & 4.4 & T & T & F & - & - & 10346, 1520, 5426 & 13 & (17) \\ 
        \multicolumn{14}{c}{\dots} \\
        
         \hline\hline
    \end{tabular}
    }
    }\vspace{0.25cm}
    {\small (*) \textbf{References:} (1): \citet{Bonaca2019}, (2): \citet{Caldwell2020}, (3): \citet{Chandra2022}, (4): \citet{Ferguson2022}, (5):~\citet{Grillmair2019}, (6): \citet{Grillmair2022}, (7): \citetalias{Ibata2021}, (8): \citet{Koposov2019}, (9): \citet{Koposov2023}, (10): \citet{Li2021}, (11): \citet{Malhan2021b}, (12): \citet{Palau2019}, (13): \cite{Palau2021},  (14): \citet{Price-Whelan2018}, (15): \citet{Price-Whelan2019}, (16):~\citetalias{Shipp2019}, (17): \citet{Thomas2022}, (18): \citet{Williams2011}, (19): \citet{Yang2023}, (20):~\citet{Yuan2022}.}
\end{table*}

\section{Membership probability of RRL}

In Table \ref{tb:membs} we show all the RRL with $p_{\text{memb}}>0.5$ and their properties: distance, period, amplitude, type, and photometric metallicity. The RRL that are identified as members are tagged as \texttt{Member detection = True}. The full table can be found in the online version.

\begin{table*}[h!]
    \caption{First 5 rows of the RRL with a probability higher than 50\% of belonging to one of the studied tracks. If the star is closer than $3\sigma_{\phi_2}$ from the sky track then is considered a member of the stream (Member detection = T), if not it is used to estimate the number of contaminants. The metallicity provenance indicate if the metallicity of the RRL was obtained from \citet[][Li23]{Li2023} or drawn from the metallicity distribution of the halo (Halo run). The period and amplitude uncertainties are $\sim5\times10^{-6}$ days and $\sim0.02$~mag, respectively.} \label{tb:membs}
    \centerline{
    \resizebox{1.05\textwidth}{!}{%
    \small{
    \begin{tabular}{cccccccccc}
         \hline\hline
         Gaia DR3 & $d$ & Period & Amp.-V & Type & [Fe/H] & [Fe/H]& $P_{\text{memb}}$ & Member & TrackName \\
         \texttt{source\_id} & (kpc) & (d) & (mag) & & adopted &provenance & & detection\\
         
         \hline\\
            5380847251631779840 & 18.3 $\pm$ 1.3 & 0.381581 & 0.43  & RRc & -2.08 $\pm$ 0.21 & Li23 & 1.00 & True & Orphan-K19 \\
            5385327207462696576 & 16.9 $\pm$ 1.1 & 0.294440 & 0.41 & RRc & -1.68 $\pm$ 0.20 & Li23 & 1.00 & True & Orphan-K19 \\
            5379867376317392384 & 16.9 $\pm$ 1.1 & 0.544678 & 0.75 & RRab & -1.63 $\pm$ 0.35 & Li23 & 1.00 & True & Orphan-K19 \\
            6456012182978611072 & 31.9 $\pm$ 2.1 & 0.378605 & 0.42 & RRc & -1.94 $\pm$ 0.22 & Li23 & 1.00 & True & Orphan-K19 \\
            6463963644911988864 & 37.1 $\pm$ 2.8 & 0.379038 & 0.53 & RRc & -2.53 $\pm$ 0.21 & Li23 & 1.00 & False & Orphan-K19 \\
            
        \multicolumn{10}{c}{\dots} \\
        \hline\hline
    \end{tabular}
    }}}
\end{table*}

\end{appendix}

\end{document}